\documentclass[aps,prb,twocolumn,groupedaddress]{revtex4-1}

\usepackage{relsize}
\usepackage{mathtools}
\usepackage{amssymb}
\usepackage{amsmath}
\usepackage{braket}
\usepackage{pgf}

\newcommand{\eM}     {$\epsilon$-machine}
\newcommand{\eMs}    {$\epsilon$-machines}

\newcommand{\Cmu}		{ {C_\mu}}

\newcommand{\eMSR}    {{{$\epsilon$}MSR}}

\newcommand{\Hagg} {H\"{a}gg}

\newcommand{\TransMatHagg} {{\sf{T}}}
\newcommand{\TransMatABC} {{\mathcal{T}}}
\newcommand{\IdentMat} {{\mathbb{I}}}
\newcommand{\abcT} {\TransMatABC} 
\newcommand{\HaggT} {\TransMatHagg}
\newcommand{\HaggTzero} {\HaggT^{[0]}} 
\newcommand{\HaggTone} {\HaggT^{[1]}} 
\newcommand{\abcTA} {\abcT^{[A]}} 
\newcommand{\abcTB} {\abcT^{[B]}} 
\newcommand{\abcTC} {\abcT^{[C]}} 
\newcommand{\TransMatSet} {{\mathcal{\mathbf{T}}}}
\newcommand{\TransMat} {{\mathcal{T}}}
\newcommand{\w} {{w}}
\newcommand{\symbolABC} {x} %{{\sigma}}  % consistent w/ $\ms$ command from cmmechabbrev
\newcommand{\symbolHagg} {{s}}
\newcommand{\StateABC} {{{\mathcal S}}}
\newcommand{\StateHagg} {{{\sf S}}}
\newcommand{\Machine} {{\Gamma}}
\newcommand{\Alphabet} {{\mathcal A}} 
\newcommand{\States} {{\mathbb{S}}}

\newcommand{\AlphabetABC} {{{\mathcal A}_{\rm P}}}
\newcommand{\AlphabetHagg} {{{\mathcal A}_{\rm H}}}
\newcommand{\MachineABC} {{$ABC$-machine}}
\newcommand{\MachineHagg} {{H\"{a}gg-machine}}
\newcommand{\Dist} {{\boldsymbol{\pi}}}
\newcommand{\DistOne} {\ket{\boldsymbol{1}}}
\newcommand{\DistOneBackward} {{\bra{\boldsymbol{1}}}}
\newcommand{\NstatesABC} {{M}_{\rm P}}
\newcommand{\NstatesHagg} {{\sf M}_{\rm H}}
\newcommand{\coperator} {{\hat{{\rm {c}}}}}
\newcommand{\aoperator} {{\hat{{\rm {a}}}}}
\newcommand{\soperator} {{\hat{{\rm {s}}}}}
\newcommand{\xoperator} {{\hat{\xi}}}

\newcommand{\eg} {{\it {e.g.}}}
\newcommand{\ie} {{\it {i.e.}}}
\newcommand{\etal} {{\it {et al.}}}
\newcommand{\Qxin} {{Q_{\xi}(n)}}
\newcommand{\Qcn}  {{Q_{\rm {c}}(n)}}
\newcommand{\Qan}  {{Q_{\rm {a}}(n)}}
\newcommand{\Qsn}  {{Q_{\rm {s}}(n)}}

\newcommand{\QcThree} {{Q_{\rm c}(3)}}
\newcommand{\QsOne} {{Q_{\rm s}(1)}}

\newcommand{\OneHalf} {{\frac{1}{2}}}
\newcommand{\CFBraKet}{\Braket{\abcT_\lambda ^{\xi(\Alphabet)}}}
\newcommand{\CFBraKeto}{\Braket{\abcT_{\lambda, 0} ^{\xi(\Alphabet)}}}
\newcommand{\CFBraKetm}{\Braket{\abcT_{\lambda, m} ^{\xi(\Alphabet)}}}
\DeclareMathOperator*{\argmax}{arg\,max}

\begin{document}

%\preprint{}

\title{Pairwise Correlations in Layered Close-Packed Structures}

\author{{P.~M.~Riechers}}
\email[]{pmriechers@ucdavis.edu}
\affiliation{{Complexity Sciences Center \& Physics Department,
	University of California, One Shields Avenue,
	{Davis, California} 95616, {USA}}}

\author{D.~P.~Varn}
\email[]{dpv@complexmatter.org}
\homepage[]{http://wissenplatz.org}
\affiliation{{Complexity Sciences Center \& Physics Department,
	University of California, One Shields Avenue,
	{Davis, California} 95616, {USA}}}

\author{J.~P.~Crutchfield}
\email[]{chaos@ucdavis.edu}
\homepage[]{http://csc.ucdavis.edu/$\sim$chaos/}
\affiliation{{Complexity Sciences Center \& Physics Department,
	University of California, One Shields Avenue,
	{Davis, California} 95616, {USA}}}
\affiliation{{Santa Fe Institute, 1399 Hyde Park Road, {Santa Fe, New Mexico}
	87501, {USA}}}

\date{\today}

\begin{abstract}
Given a description of the stacking statistics of layered close-packed
structures in the form of a hidden Markov model, we develop analytical
expressions for the pairwise correlation functions between the layers. These
may be calculated analytically as explicit functions of model parameters or the
expressions may be used as a fast, accurate, and efficient way to obtain
numerical values. We present several examples, finding agreement with previous
work as well as deriving new relations.
\end{abstract}

\pacs{Put Pacs codes here}

\maketitle
\section{Introduction}

There has long been an interest in planar defects or stacking faults (SFs) in
crystals~\cite{Hend42a,Wils42a}. With the recent realization of the
technological import of many materials prone to
SFs---graphene~\cite{Neto09a,Geim13a} and SiC~\cite{Zeke11a} being but two
well known examples---that interest has only grown. Since SFs shift an entire
layer of atoms, it is not unexpected that they can profoundly affect material
properties. Many of these materials have more than one stable stacking
configuration and additionally many metastable ones can exist~\cite{Seba94a},
as well as many stacking configurations that show varying amounts of disorder.
Thus, understanding SFs, perhaps as a prelude to controlling their kind and
placement, presents a significant, but compelling challenge.

Disordered layered materials are often studied via the pairwise correlations
between layers, as these correlations are experimentally accessible from the
Fourier analysis of a diffraction pattern
(DP)~\cite{Este01a,Este01b,Varn02a,Varn13a} or directly from simulation
studies~\cite{Kabr88a,Kabr95a,Shre96a,Shre96b,Shre96c,Shre97a,Varn04a}.  Such
studies yield important insights into the structural organization of
materials.  For example, Kabra \& Pandey~\cite{Kabr88a} were able to show
that a model of the 2H $\to$ 6H{\footnote{We will use the Ramsdell
notation~\cite{Orti13a,Seba94a} to describe well known crystalline stacking
structures.}} transformation in SiC could retain long-range order even as
the short-range order was reduced. Tiwary \& Pandey~\cite{Tiwa07a}
calculated the size of domains in a model of randomly faulted close-packed
structure (CPSs) by calculating the (exponential) decay rate of pairwise
correlation functions between layers. Recently Estevez-Rams
\etal~\cite{Este08a} derived analytical expressions for the correlation
functions for CPSs that contained both random growth and deformation faults,
and Beyerlein \etal~\cite{Beye11a} demonstrated that correlation functions
in finite-sized FCC crystals depend not only on the kind and amount of
faulting, but additionally on their placement.

Beyond the study of layered materials, pairwise correlation information, in the
form of {\emph {pair distribution functions}} (PDFs), has recently attracted
significant attention~\cite{Egam13a}. However, as useful as the study of
pairwise correlation information is, it does \emph{{not}} provide a complete
description of the specimen. Indeed, it has long been known that very different
atomic arrangements of atoms can reproduce the same PDF, although there has
been recent progress in reducing this degeneracy~\cite{Clif10a}. Nor are they
in general suitable for calculating material properties, such as conductivities
or compressibilities.

For crystalline materials, a complete description of the specimen comes in the
form of its crystal structure, \ie, the specification of the placement of all
the atoms within the unit cell, as well as the description of how the unit
cells are spatially related to each other, commonly referred to as the lattice.
Determining these quantities for specimens and materials is of course the
traditional purview of crystallography. For disordered materials, a similar
formalism is required that provides a unified platform not only to calculate
physical quantities of interest but also to give insight into their physical
structure. For layered materials, where there is but one axis of interest,
namely the organization along the stacking direction, such a formalism has been
identified~\cite{Varn02a,Varn07a,Varn13a}, and that formalism is {\em
{computational mechanics}}~\cite{Crut89a,Crut12a}. The mathematical entity
that gives a compact, statistical description of the disordered material (along
its stacking direction) is its \emph{\eM}, a kind of hidden Markov model
(HMM)~\cite{Rabi89a,Elli95a}. Computational mechanics also has the advantage
of encompassing traditional crystal structures, so both ordered and disordered
materials can be treated on the same footing in the same formalism.

It is our contention that an \eM\ describing a specimen's stacking includes all
of the structural information necessary to calculate physical quantities that
depend on the stacking statistics~\cite{Varn14a}. In the following, we
demonstrate how pairwise correlation functions can be either calculated {\em
analytically} or to a high degree of numerical certainty for an arbitrary
HMM and, thus, for an arbitrary \eM. Previous researchers often calculated pairwise correlation functions for particular realizations of stacking
configurations~\cite{Berl86a,Kabr88a,Shre97a,Este01a} or from analytic
expressions constructed for particular models~\cite{Tiwa07a,Este08a}. The
techniques developed here, however, are the first generally applicable methods
that do not rely on samples of a stacking sequence. The result delivers both an
analytical solution and an efficient numerical tool. And while we will
specialize to the case of CPSs for concreteness, the methods developed are
extendable to other materials and stacking geometries. 

Our development is organized as follows: In \S \ref{Stacking Notations} we
introduce nomenclature. In \S \ref{sec:HaggToABC} we develop an algorithm to
change between different representations of stacking sequences. In \S
\ref{sec:CFsFromHMMs} we derive expressions, our main results, for the pairwise
correlation functions between layers in layered CPSs. In \S \ref{Examples} we
consider several examples; namely, (i) a simple stacking process that
represents the 3C crystal structure or a completely random stacking depending
on the parameter choice, (ii) a stacking process that represents random growth
and deformation faults, and (iii) a stacking process inspired by recent
experiments in 6H-SiC. And, in \S \ref{Conclusions} we give our conclusions and
directions for future work.

\section{Definitions and Notations}
\label{Stacking Notations}

We suppose the layered material is built up from identical sheets called {\em
{modular layers}} (MLs)~\cite{Pric83a,Ferr08a}. The MLs are completely ordered
in two dimensions and assume only one of three discrete positions, labeled $A$,
$B$, or $C$~\cite{Ashc76a,Seba94a}. These represent the physical placement of
each ML and are commonly known as the $ABC$-notation~\cite{Orti13a}. We define the set of possible
orientations in the $ABC$-notation as $\AlphabetABC = \{A, B, C\}$. We further
assume that the MLs obey the same stacking rules as CPSs, namely that two
adjacent layers may not have the same orientation; \ie, stacking sequences
$AA$, $BB$ and $CC$ are not allowed. Exploiting this constraint, the stacking
structure can be represented more compactly in the \Hagg-notation: one takes
the transitions between MLs as being either cyclic, ($A \to B, B \to C,$ or $C
\to A$), and denoted as `+'; or anticyclic, ($A \to C, C \to B,$ or $B \to A$),
and denoted as `-'. The \Hagg-notation then gives the relative orientation of
each ML to its predecessor. It is convenient to identify the usual
\Hagg-notation `+' as `1' and `-' as `0'. Doing so, we define the set of
possible relative orientations in the \Hagg-notation as $\AlphabetHagg =
\{0,1\}$.  These two notations---$ABC$ and \Hagg---carry an identical message,
up to an overall rotation of the specimen. Alternatively, one can say that there is
freedom of choice in labeling the first ML.

\subsection{Correlation functions}
\label{CorrelationFunctions} 

Let us define three
statistical quantities, $\Qcn$, $\Qan$, and $\Qsn$~\cite{Yi96a}: the
pairwise \emph{correlation functions} (CFs) between MLs, where c, a, and
s stand for \emph{cyclic}, \emph{anticyclic}, and \emph{same},
respectively.  $\Qcn$ is the probability that any two MLs at a separation of
$n$ are cyclically related. $\Qan$ and $\Qsn$ are defined in a similar
fashion.\footnote{As yet, there is no consensus on notation for these
quantities. Warren~\cite{Warr69a} uses $P_m^0, P_m^+$, and $P_m^-$, Kabra
\& Pandey~\cite{Kabr88a} call these $P(m), Q(m)$, and $R(m)$, and
Estevez~\etal~\cite{Este08a} use $P_0(\Delta), P_{\rm f}(\Delta)$, and $P_{\rm
b}(\Delta)$. Since we prefer to reserve the symbol `$P$' for other
probabilities previously established in the literature, here and elsewhere we
follow the notation of Yi \& Canright~\cite{Yi96a}, with a slight
modification of replacing `$Q_{\rm{r}}(n)$' with `$\Qan$'.} Since these are probabilities: $0
\leq \Qxin \leq 1$, where $\xi \in \{\rm{c,a,s} \}$. Additionally, at each $n$ it is
clear that $\sum_{\xi} \Qxin = 1$. Notice that the CFs are defined in terms of
the $ABC$-notation.

\subsection{Representing layer stacking as a hidden process}
\label{TransitionMatrices}

We chose to represent a stacking sequence as the output of discrete-step,
discrete-state hidden Markov model (HMM). A HMM $\Gamma$ is an ordered tuple
$\Machine = (\Alphabet, \States, \mu_0, \TransMatSet)$, where $\Alphabet$ is
the set of symbols that one observes as the HMM's output, often called an
alphabet, $\States$ is a finite set of $M$ internal states, $\mu_0$ is an
initial state probability distribution, and $\TransMatSet$ is a set of matrices
that give the probability of making a transition between the states while
outputting one of the symbols in $\Alphabet$. These transition probability
matrices or more simply \emph{transition matrices} (TMs)~\cite{Paz71a,Karl75a}
are usually written:
\begin{align}
  \TransMat^{[\symbolHagg]} =
  \begin{bmatrix}  \nonumber
  \Pr(\symbolHagg, \StateHagg_1|\StateHagg_1 ) & \Pr(\symbolHagg, \StateHagg_2|\StateHagg_1 ) & \cdots & \Pr(\symbolHagg, \StateHagg_M|\StateHagg_1 ) \\
  \Pr(\symbolHagg, \StateHagg_1|\StateHagg_2 ) & \Pr(\symbolHagg, \StateHagg_2|\StateHagg_2 ) & \cdots & \Pr(\symbolHagg, \StateHagg_M|\StateHagg_2 ) \\
  \vdots  & \vdots  & \ddots & \vdots  \\
  \Pr(\symbolHagg, \StateHagg_1|\StateHagg_M ) & \Pr(\symbolHagg, \StateHagg_2|\StateHagg_M ) & \cdots & \Pr(\symbolHagg, \StateHagg_M|\StateHagg_M ) \\
  \end{bmatrix} ,
\end{align}
where $\symbolHagg \in \Alphabet$ and $\StateHagg_1, \StateHagg_2, \dots, \StateHagg_M \in \States$.

For a number of purposes it is convenient to work directly with the internal
state TM, denote it $\TransMat$. This is the matrix of state transition
probabilities regardless of symbol, given by the sum of the symbol-labeled TMs:
$\TransMat = \sum_{x \in \Alphabet} \TransMat^{[x]}$. For example, the
internal state distribution evolves according to $\bra{\boldsymbol{\mu_1}} = 
\bra{\boldsymbol{\mu_0}} \TransMat$.
Or, more generally, $\bra{\boldsymbol{\mu_L}} = \bra{\boldsymbol{\mu_0}} \TransMat^L$. 
(In this notation, state distributions are row vectors.) 
In another use, one finds the stationary state probability distribution:
\begin{align}
    \bra \Dist 
    &= \begin{bmatrix}    \nonumber
        \Pr({{\StateHagg}_1}) & \Pr({{\StateHagg}_2}) &  \dotsm & 
        \Pr({{\StateHagg}_{M}})
    \end{bmatrix}, 
\end{align}
as the left eigenvector of $\TransMat$ normalized in probability:
\begin{align}
   \bra \Dist = \bra \Dist \TransMat ~.
    \label{eq:ProbabilityDistribution}
\end{align}  

The probability of any finite-length sequence of symbols can be computed exactly from these
objects using linear algebra.
In particular, a length-$L$ `word' 
$w = \symbolHagg_0 \symbolHagg_1 \dots \symbolHagg_{L-1} \in \Alphabet^L$, 
where $\Alphabet^L$ is the set of length-$L$ sequences, has the stationary probability:
\begin{align*}
\Pr(w) &= \bra \Dist \TransMat^{[w]} \DistOne \\ 
& = \bra \Dist \TransMat^{[\symbolHagg_0]} \TransMat^{[\symbolHagg_1]} \dotsm \TransMat^{[\symbolHagg_{L-1}]} \DistOne~,
\end{align*}
where $\DistOne$ is the column-vector of all ones. 

As a useful convention, we will use bras $\bra{\cdot}$ to denote row vectors
and kets $\ket{\cdot}$ to denote column vectors. On the one hand, any object
closed by a bra on the left and ket on the right is a scalar and commutes as a
unit with anything. On the other hand, a ket--bra $\ket{\cdot} \bra{\cdot}$ has
the dimensions of a square matrix.

To help make these ideas concrete, let us consider a CPS stacked according to the \emph{Golden Mean
Process} (GMP) represented in the \Hagg-notation. This process has been
previously treated in the context of CPSs in Varn~\cite{Varn13b}. Any
sequence is allowed as long as there are no consecutive 0s. This is
accomplished by examining the previous observed symbol: if it is 1, then the
next symbol in the sequence is either 0 or 1 with equal probability; if it is
0, then the next symbol is necessarily 1. Thus, there are two states in the
\Hagg-machine, corresponding to the above two conditions. Let us call these
states $\mathcal{U}$ (next symbol is a 0 or 1 with equal probability) and
$\mathcal{V}$ (next symbol is a 1). And so, we say $\States
=\{\mathcal{U},\mathcal{V}\}$. The two 2-by-2 TMs for this process---one for
each symbol in the alphabet---are given by:
\begin{align*}
  \begin{array}{r@{\mskip\thickmuskip}l}
       \TransMatHagg^{[0]}  
           &=
              \begin{bmatrix}
              0 & \OneHalf \\
              0 & 0
              \end{bmatrix} 
  \end{array} 
  {\rm {and}}
  \begin{array}{r@{\mskip\thickmuskip}l}
            \TransMatHagg^{[1]} 
     &=
         \begin{bmatrix}
         \OneHalf & 0 \\
         1 & 0
        \end{bmatrix}~.        
          \end{array}
\end{align*}
The GMP has internal-state TM:
\begin{align}
\TransMatHagg 
&= \TransMatHagg^{[0]} + \TransMatHagg^{[1]}    \nonumber \\ 
&=
\begin{bmatrix} \nonumber
\OneHalf & \OneHalf \\
1 & 0
\end{bmatrix}~.
\end{align}
The asymptotic state probabilities are given by
$\bra{\Dist_{\textrm{H}}} = \Big[ \tfrac{2}{3} ~~ \tfrac{1}{3} \Big]$.

\begin{figure}
\begin{center}
\includegraphics[width=0.35\textwidth]{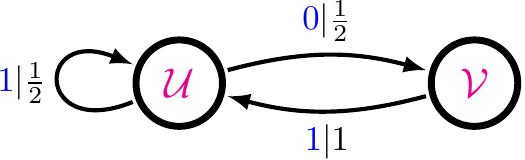}
\end{center}
\caption{The GM Process written as a \MachineHagg. The circles indicate states,
  and the arcs between them are transitions, labeled by $s|p$, where $s$ is the
  symbol emitted upon transition and $p$ is the probability of making such a
  transition.
  }
\label{FigGMProcessHagg}
\end{figure}

In this way, the \MachineHagg\ for the GM Process is defined as
$\Machine^{({\rm H})}_{\rm {GM}} = (\Alphabet, \States, \mu_0,
\TransMatSet) = (\{0,1\}, \{\mathcal{U},\mathcal{V}\}, 
\left[ \tfrac{2}{3} ~~ \tfrac{1}{3} \right] , \{\TransMatHagg^{[0]}, 
\TransMatHagg^{[1]}\})$.\footnote{Here and in the examples of \S\ref{Examples},
we take the stationary state probability distribution $\Dist$ as the initial probability state distribution $\mu_0$,
as we are interested for now in the the long term behavior.}
HMMs are often conveniently depicted using labeled directed graphs.
As an example, the GM Process's HMM is shown in Fig.~\ref{FigGMProcessHagg}.  

\section{Expanding the \MachineHagg\ to the \MachineABC}
\label{sec:HaggToABC}

While simulation studies~\cite{Varn04a} and $\epsilon$-Machine Spectral
Reconstruction (\eMSR)~\cite{Varn02a,Varn07a,Varn13a,Varn13b} express stacking
structure in terms of the \MachineHagg, for some calculations it is more
convenient to represent the stacking process in terms of the \MachineABC. Here,
we give a graphical procedure for expanding the \MachineHagg\ into the
\MachineABC\ and then provide an algebraically equivalent algorithm.  We note
that this expansion procedure is \emph{not} unique and can vary up to an
overall rearrangement of the columns and rows of the resulting \MachineABC\ TM.
This difference, of course, does not alter the results of calculations of
physical quantities.

\subsection{Graphical expansion method}
\label{GraphicalExpansion}

\begin{figure}
\begin{center}
\includegraphics[width=0.22\textwidth]{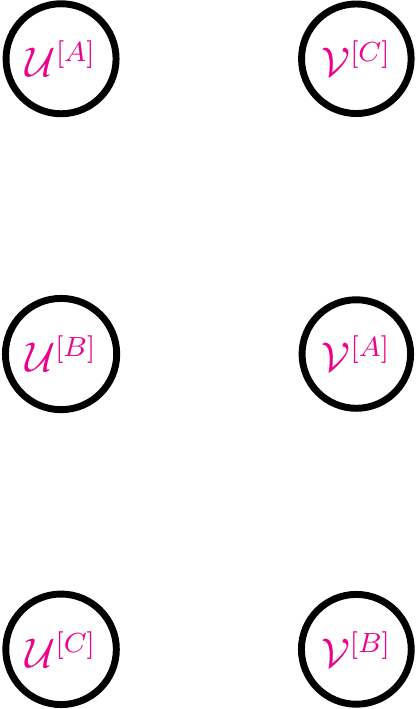}
\end{center}
\caption{The first step in expanding a \MachineHagg\ into a \MachineABC\ is to
  treble the number of states.
  }
\label{FigGMProcessABC1}
\end{figure}

Recall that the \Hagg-notation and the $ABC$-notation are equivalent 
representations of the stacking structure,
up to an overall rotation of the crystal. Stated alternatively, in
the \Hagg-notation, there is an ambiguity concerning the orientation of each
ML---it could be either $A$, $B$ or $C$.  To account for this degeneracy, when
we transform to the $ABC$ representation, we triple the size of the
\MachineHagg.  As a first step, one writes down three states for each state
found in the \MachineHagg, but not the transitions between them.  To
distinguish among these new states of the triplet, label each with a superscript
($A$, $B$ or $C$) indicating the last ML added to arrive at that state.
Symbolically, this is stated:
\begin{align}
\left\{\StateHagg_i \right\} 
\xRightarrow{\text{\Hagg\ to } ABC} 
\left\{ \StateABC_i^{[A]}, \text{ } \StateABC_i^{[B]}, \text{ } \StateABC_i^{[C]} \right\}~. \nonumber
\end{align}
Transitions between states on the \MachineABC\ still respect the same state
labeling scheme as on the \MachineHagg\ (explained below), but now they store the ML information.
Transitions between states on the \MachineHagg\ that were labeled with $1$
advance the stacking sequence cyclically (\ie, $A \to B \to C \to A$) and the
corresponding transitions on the \MachineABC\ reflect this by taking the ML
label on the initial state and advancing it cyclically. In a completely
analogous way, transitions labeled 0 on the \MachineHagg\ advance the states on the 
\MachineABC\ in an anticyclical fashion (\ie, $A \to C \to B \to A$).

Continuing our GM Process example, let us write out the six ($= 3 \times 2$)
states labeled with superscripts to distinguish them. This is done in
Fig.~\ref{FigGMProcessABC1}. (It does not matter in what order these states are
labeled. The scheme chosen in Fig.~\ref{FigGMProcessABC1} turns out to be
convenient given the state-to-state transition structure of the final
\MachineABC, but any arrangement is satisfactory.) The transitions between the
states on the \MachineABC\ preserve the labeling scheme of the original
\MachineHagg. That is, if in the original \MachineHagg\ there is transition
$\StateHagg_i \xrightarrow{\symbolHagg|p} \StateHagg_j$, then there must be
\emph{three} similar transitions on the \MachineABC\ of the form
$\StateABC_i^{[\symbolABC]} \xrightarrow{\symbolABC^{\prime}|p}
\StateABC_j^{[\symbolABC^{\prime}]}$, with $\symbolABC, \symbolABC^{\prime} \in
\{A,B,C\}$. Additionally, the transitions on the \MachineABC\ corresponding to
the transitions on the \MachineHagg\ have the same probability.

\begin{figure}
\begin{center}
\includegraphics[width=0.25\textwidth]{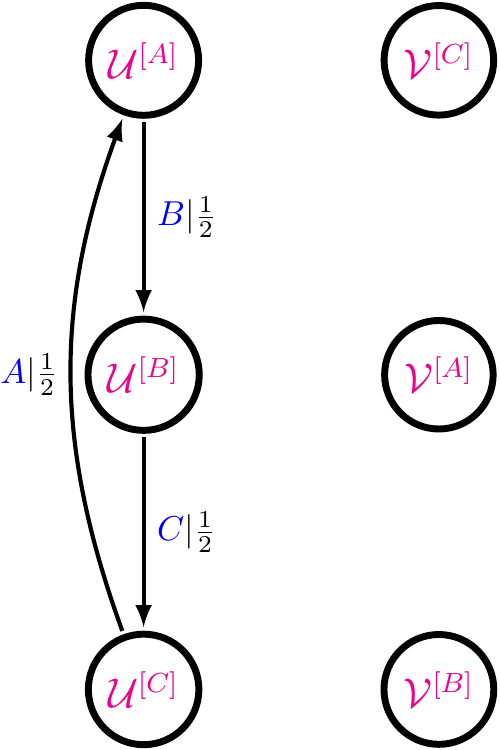}
\end{center}
\caption{The second step in expanding the \MachineHagg\ into the \MachineABC\ is to 
             add the transitions. Here, a single transition on the example \MachineHagg, 
             $\mathcal{U} \xrightarrow{1|\OneHalf} \mathcal{U}$, is expanded into {\emph {three}}
             transitions on the \MachineABC.}
\label{FigGMProcessABC2}
\end{figure}

\begin{figure}
\begin{center}
\includegraphics[width=0.24\textwidth]{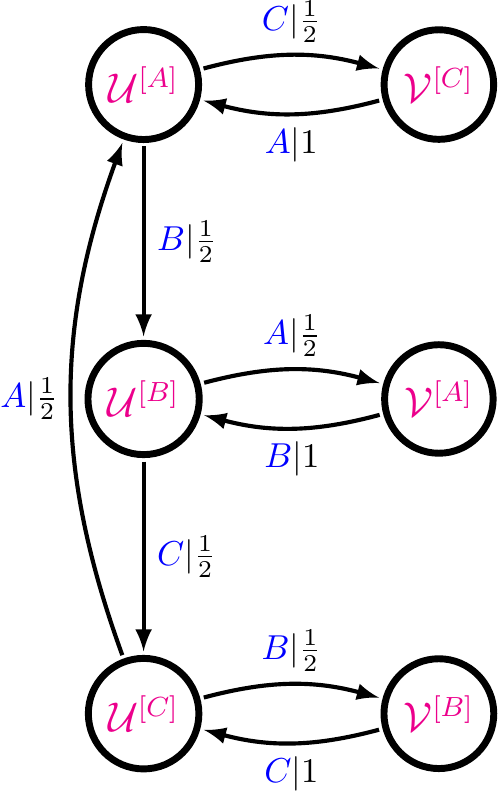}
\end{center}
\caption{The completely expanded six-state \MachineABC\ that corresponds to the two-state 
              \MachineHagg\ shown in Fig.~\ref{FigGMProcessHagg}.}
\label{FigGMProcessABC3}
\end{figure}

Let us consider the self-state transition on the \MachineHagg\ shown in
Fig.~\ref{FigGMProcessHagg}: $\mathcal{U} \xrightarrow{1|\OneHalf}
\mathcal{U}$.  Since the corresponding transitions on the \MachineABC\ still
respect the state labeling scheme, the self-loop on $\mathcal{U}$ only induces transitions among the
{${\mathcal{U}}^{[\symbolABC]}$}. Since a $1$ advances the stacking sequence cyclically, the appropriate transitions are:
\begin{align}
{\mathcal{U}}^{[A]} & \xrightarrow{B|\OneHalf} {\mathcal{U}}^{[B]}, \nonumber \\
{\mathcal{U}}^{[B]} & \xrightarrow{C|\OneHalf} {\mathcal{U}}^{[C]}, \nonumber \\
{\mathcal{U}}^{[C]} & \xrightarrow{A|\OneHalf} {\mathcal{U}}^{[A]} ~. \nonumber 
\end{align}
This is illustrated in Fig.~\ref{FigGMProcessABC2}. Applying the same procedure to the other
transitions on the \MachineHagg, \ie, $\mathcal{U} \xrightarrow{0|\OneHalf} \mathcal{V}$
and $\mathcal{V} \xrightarrow{1|1} \mathcal{U}$, 
results in the completely expanded \MachineABC,
and this is shown in Fig.~\ref{FigGMProcessABC3}. 

We are now able to write down the stacking process for the GM Process from its
expanded graph, Fig.~\ref{FigGMProcessABC3}. First, we note that the alphabet
is ternary: $\AlphabetABC = \{A, B, C\}$. Second, there are six states on the
\MachineABC, \ie, $\StateABC = \{ {\mathcal{U}}^{[A]}, {\mathcal{U}}^{[B]},
{\mathcal{U}}^{[C]}, {\mathcal{V}}^{[A]}, {\mathcal{V}}^{[B]},
{\mathcal{V}}^{[C]} \}$.  Ordering the states as above, the TMs may be directly
constructed from the expanded graph, and are given by:
\begin{align}
\TransMatABC^{[A]} &=
\begin{bmatrix}  \nonumber
0 & 0 & 0 & 0 & 0 & 0   \\
0 & 0 & 0 & \OneHalf & 0 & 0   \\
\OneHalf & 0 & 0 & 0 & 0 & 0   \\
0 & 0 & 0 & 0 & 0 & 0   \\
0 & 0 & 0 & 0 & 0 & 0   \\
1 & 0 & 0 & 0 & 0 & 0  
\end{bmatrix}, 
\TransMatABC^{[B]} =
\begin{bmatrix}  \nonumber 
0 & \OneHalf & 0 & 0 & 0 & 0   \\
0 & 0 & 0 & 0 & 0 & 0   \\
0 & 0 & 0 & 0 & \OneHalf & 0   \\
0 & 1 & 0 & 0 & 0 & 0   \\
0 & 0 & 0 & 0 & 0 & 0   \\
0 & 0 & 0 & 0 & 0 & 0  
\end{bmatrix},
\end{align}
and
\begin{align}
\TransMatABC^{[C]} 
&=
\begin{bmatrix}  \nonumber
0 & 0 & 0 & 0 & 0 & \OneHalf   \\
0 & 0 & \OneHalf & 0 & 0 & 0   \\
0 & 0 & 0 & 0 & 0 & 0   \\
0 & 0 & 0 & 0 & 0 & 0   \\
0 & 0 & 1 & 0 & 0 & 0   \\
0 & 0 & 0 & 0 & 0 & 0  
\end{bmatrix}~. 
\end{align}
As before, the internal-state TM is simply the sum of the symbol-specific TMs,
given by $\TransMatABC = \TransMatABC^{[A]} +\TransMatABC^{[B]} +
\TransMatABC^{[C]}$. For the GM Process this turns out to be:
\begin{align}
\TransMatABC
&=
\begin{bmatrix}    \nonumber
0 & \OneHalf & 0 & 0 & 0 & \OneHalf   \\
0 & 0 & \OneHalf & \OneHalf & 0 & 0   \\
\OneHalf & 0 & 0 & 0 & \OneHalf & 0   \\
0 & 1 & 0 & 0 & 0 & 0   \\
0 & 0 & 1 & 0 & 0 & 0   \\
1 & 0 & 0 & 0 & 0 & 0
\end{bmatrix}~.
\end{align}

For completeness, the HMM for the GM Process in terms of the physical stacking of MLs is 
$\Machine^{({\rm P})}_{\rm {GM}} = (\Alphabet, \States, \mu_0, \TransMatSet) = \bigl( \{A,B,C\}, \bigr.$
$\{{\mathcal{U}}^{[A]}, {\mathcal{U}}^{[B]}, {\mathcal{U}}^{[C]}, 
{\mathcal{V}}^{[A]}, {\mathcal{V}}^{[B]}, {\mathcal{V}}^{[C]}\},$
$\tfrac{1}{9} \left[ 2~~2~~2~~1~~1~~1 \right],$
%\left[ \tfrac{2}{9}~~\tfrac{2}{9}~~\tfrac{2}{9}~~\tfrac{1}{9}~~\tfrac{1}{9}~~\tfrac{1}{9} \right],
$\bigl. \{\TransMatABC^{[A]}, \TransMatABC^{[B]}, \TransMatABC^{[C]}\} \bigr)$.   

\subsection{Mixing and Nonmixing State Cycles}

Observe Fig.~\ref{FigGMProcessABC3}'s directed graph is \emph{strongly
connected}---any state is accessible from any other state in a finite number of
transitions. It should be apparent that this need not have been the case. In
fact, in this example connectivity is due to the presence of the self-state
transition $\mathcal{U} \xrightarrow{1} {\mathcal{U}}$. The latter guarantees a
strongly connected expanded graph. Had this transition been absent on the
\MachineHagg, such that there were only transitions of the form $\mathcal{U}
\xrightarrow{0} \mathcal{V}$ and $\mathcal{V} \xrightarrow{1} \mathcal{U}$, the
expansion would have yielded a graph with three distinct, unconnected
components. Only one of these graphs would be physically manifest. It is
sufficient to take just one component, arbitrarily assign a $A,B$ or $C$ to an
arbitrary state on that component, and then replace all of the $\{0,1\}$
transitions with the appropriate $\{A,B,C\}$ transitions, as done above.  

To determine whether the expansion process on a \MachineHagg\ results in a
strongly connected graph, one can examine the set of \emph{simple state cycles}
(SSCs) and calculate the \emph{winding number} for each.  A SSC is defined
analogous to a \emph{causal state cycle} (CSC)~\cite{Varn13a} on an \eM\ as a
``finite, closed, nonself-intersecting, symbol-specific path" along the graph.
The winding number $W$ for a SSC on a \MachineHagg\ is similar to the 
parameter $\Delta$ previously defined by Yi \& Canright~\cite{Yi96a} and the 
{\em cyclicity} ($C$)~\cite{Dorn71a} for a polytype of a CPS. $W$ differs from $C$ as the former is
not divided by the period of the cycle. We define the winding number for a SSC
as:
\begin{align}
   W^{\rm {SSC}} = n_1 - n_0~,   \nonumber
\end{align}
where $n_1$ and $n_0$ are the number of 1s and the number of $0$s encountered
traversing the SSC, respectively. We call those SSCs \emph{mixing} if $W^{\rm
{SSC}} \pmod{3} \neq 0$, and \emph{nonmixing} if $W^{\rm {SSC}} \pmod{3} = 0$.
If there is at least one mixing SSC on the \MachineHagg, then the expanded
\MachineABC\ will be strongly connected. For example, there are two SSCs on the
\MachineHagg\ for the GM Process: $[\mathcal{U}]$ and
$[\mathcal{U}\mathcal{V}]$.\footnote{We use the same nomenclature to denote a
SSC as previously used to denote a CSC: The state sequence visited traversing
the cycle is given in square brackets~\cite{Varn13a}. For those cases where an
ambiguity exists because the transition occurs on more than one symbol, we
insert a subscript in parentheses denoting that symbol.} The winding number for each is given
by $W^{[\mathcal{U}]} = 1 - 0 = 1$ and $W^{[\mathcal{U}\mathcal{V}]} = 1 - 1 =
0$.  Since $W^{[\mathcal{U}]} \neq 0$ and $[\mathcal{U}]$ is thus a mixing SSC,
the \MachineHagg\ for the GM Process will expand into a strongly connected
\MachineABC. Let us refer to those \MachineHagg s with at least one mixing SSC
as \emph{mixing \MachineHagg s} and those that do not as \emph{nonmixing
\MachineHagg s} and similarly for the corresponding \MachineABC s. We find that
mixing \MachineHagg s, and thus mixing \MachineABC s, are far more common than
nonmixing ones and that the distinction between the two can have profound
effects on the calculated quantities, such as the CFs and the
DP~\cite{Riec14c}.

\subsection{Rote expansion algorithm}
\label{AlgebraicExpansion}

To develop an algorithm for expansion, it is more convenient to change notation
slightly. Let us now denote $\StateABC$ as the set of hidden recurrent states
in the \MachineABC, indexed by integer subscripts: $\StateABC = \{\StateABC_i:
i=1, \dots, \NstatesABC\}$, where $\NstatesABC = |\StateABC|$. Define the
probability to transition from state $\StateABC_i$ to state $\StateABC_j$ on
the symbol $\symbolABC \in \AlphabetABC$ as $\TransMatABC^{[\symbolABC]}_{i,
j}$. Let's gather these state-to-state transition probabilities into a
$\NstatesABC \times \NstatesABC$ matrix, referring to it as the
\emph{$\symbolABC$-transition matrix} ($\symbolABC$-TM)
$\TransMatABC^{[\symbolABC]}$. Thus, there will be as many $\symbolABC$-TMs as
there are symbols in the alphabet of the \MachineABC, which is always
$|\AlphabetABC| = 3$ for CPSs.

As before, transitioning on symbol $1$ has a threefold degeneracy in the $ABC$
language, as it could imply any of the three transitions $(A \to B$, $B \to C$,
or $C \to A$), and similarly for 0. Thus, each labeled edge of the Hagg-machine
must be split into three distinct labeled edges of the $ABC$-machine.
Similarly, each state of the \MachineHagg\ maps onto three distinct states of
the $ABC$-machine. Although we have some flexibility in indexing states in the
resulting $ABC$-machine, we establish consistency by committing to the
following construction.\footnote{Alternative constructions merely swap the
labels of different states, but this choice of indexing affects the particular
form of the TMs and how they are extracted from the \MachineHagg\ TMs. We
choose the construction here for its intuitive and simple form.}

If $\NstatesHagg$ is the number
of states in the \MachineHagg, then $\NstatesABC = 3 \NstatesHagg$ for mixing \MachineHagg s.
(The case of nonmixing \MachineHagg s is treated afterward.) 
Let the $i^{\rm {th}}$ state of the \MachineHagg\ split into 
the ${\left( 3i-2 \right) }^{\rm {th}}$ through the ${\left( 3i \right) }^{\rm
{th}}$ states of the 
corresponding \MachineABC. Then, each labeled-edge transition from 
the $i^{\rm {th}}$ to the $j^{\rm {th}}$ states of the \MachineHagg\ maps into a 
3-by-3 submatrix for each of the three labeled TMs of the 
\MachineABC\ as:
\begin{align}
\label{eq:T0toABC}
\left\{\TransMatHagg_{ij}^{[0]} \right\} 
\xRightarrow{\text{\Hagg\ to } ABC} 
\left\{ \TransMatABC_{3i-1, 3j-2}^{[A]}, \text{ } \TransMatABC_{3i, 3j-1}^{[B]}, \text{ } \TransMatABC_{3i-2, 3j}^{[C]} \right\}
\end{align}
and
\begin{align}
\label{eq:T1toABC}
\left\{\TransMatHagg_{ij}^{[1]} \right\} 
\xRightarrow{\text{\Hagg\ to } ABC} 
\left\{ \TransMatABC_{3i, 3j-2}^{[A]}, \text{ } \TransMatABC_{3i-2, 3j-1}^{[B]}, \text{ } \TransMatABC_{3i-1, 3j}^{[C]} \right\} .
\end{align}
We can represent the mapping of Eq.\ \eqref{eq:T0toABC} and Eq.\ \eqref{eq:T1toABC} more 
visually with the following equivalent set of statements:
%\vspace{-0.1in}
\begin{align}
\begin{pmatrix}
\TransMatABC_{3i-2, 3j-2}^{[A]} & \TransMatABC_{3i-2, 3j-1}^{[A]} & \TransMatABC_{3i-2, 3j}^{[A]} \\
\TransMatABC_{3i-1, 3j-2}^{[A]} & \TransMatABC_{3i-1, 3j-1}^{[A]} & \TransMatABC_{3i-1, 3j}^{[A]} \\
\TransMatABC_{3i, 3j-2}^{[A]}    & \TransMatABC_{3i, 3j-1}^{[A]}    & \TransMatABC_{3i, 3j}^{[A]}
\end{pmatrix}
=
\begin{pmatrix}
0 & 0 & 0 \\
\TransMatHagg_{ij}^{[0]} & 0 & 0 \\
\TransMatHagg_{ij}^{[1]} & 0 & 0
\end{pmatrix} ,
\end{align}
\vspace{-0.1in}
\begin{align}
\begin{pmatrix}
\TransMatABC_{3i-2, 3j-2}^{[B]} & \TransMatABC_{3i-2, 3j-1}^{[B]} & \TransMatABC_{3i-2, 3j}^{[B]} \\
\TransMatABC_{3i-1, 3j-2}^{[B]} & \TransMatABC_{3i-1, 3j-1}^{[B]} & \TransMatABC_{3i-1, 3j}^{[B]} \\
\TransMatABC_{3i, 3j-2}^{[B]}    & \TransMatABC_{3i, 3j-1}^{[B]} &    \TransMatABC_{3i, 3j}^{[B]}
\end{pmatrix}
=
\begin{pmatrix}
0 & \TransMatHagg_{ij}^{[1]} & 0 \\
0 & 0 & 0 \\
0 & \TransMatHagg_{ij}^{[0]} & 0
\end{pmatrix} ,
\end{align}
\vspace{-0.1in}
and:
%\vspace{-0.1in}
\begin{align}
\begin{pmatrix}
\TransMatABC_{3i-2, 3j-2}^{[C]} & \TransMatABC_{3i-2, 3j-1}^{[C]} & \TransMatABC_{3i-2, 3j}^{[C]} \\
\TransMatABC_{3i-1, 3j-2}^{[C]} & \TransMatABC_{3i-1, 3j-1}^{[C]} & \TransMatABC_{3i-1, 3j}^{[C]} \\
\TransMatABC_{3i, 3j-2}^{[C]}    & \TransMatABC_{3i, 3j-1}^{[C]}    & \TransMatABC_{3i, 3j}^{[C]}
\end{pmatrix}
=
\begin{pmatrix}
0 & 0 & \TransMatHagg_{ij}^{[0]} \\
0 & 0 & \TransMatHagg_{ij}^{[1]} \\
0 & 0 & 0
\end{pmatrix} ,
\end{align}
which also yields the 3-by-3 submatrix for the \emph{unlabeled} $ABC$ TM in
terms of the \emph{labeled H\"{a}gg TMs}:
\begin{align}
\begin{pmatrix}
\TransMatABC_{3i-2, 3j-2} & \TransMatABC_{3i-2, 3j-1} & \TransMatABC_{3i-2, 3j} \\
\TransMatABC_{3i-1, 3j-2} & \TransMatABC_{3i-1, 3j-1} & \TransMatABC_{3i-1, 3j} \\
\TransMatABC_{3i, 3j-2}    & \TransMatABC_{3i, 3j-1}    & \TransMatABC_{3i, 3j}
\end{pmatrix}
=
\begin{pmatrix}
0 & \TransMatHagg_{ij}^{[1]} & \TransMatHagg_{ij}^{[0]} \\
\TransMatHagg_{ij}^{[0]} & 0 & \TransMatHagg_{ij}^{[1]} \\
\TransMatHagg_{ij}^{[1]} & \TransMatHagg_{ij}^{[0]} & 0
\end{pmatrix} .
\end{align}

Furthermore, for mixing \MachineHagg s, the probability from the stationary
distribution over their states maps to a triplet of probabilities for the
stationary distribution over the \MachineABC\ states: 
\begin{align}
\left\{ {p^{\text{H}}}_{i} \right\} 
\xRightarrow{\text{\Hagg\ to } ABC} 
\left\{ 3p_{3i-2}, \text{ } 3p_{3i-1}, \text{ } 3p_{3i} \right\}
\end{align}
such that:
\begin{align}
{\bra \Dist } &= \;
  \begin{bmatrix} p_1 & p_2 & p_3 & p_4 & \dots & p_{\NstatesABC - 1} & p_{\NstatesABC} \end{bmatrix} \nonumber \\
  & =\tfrac{1}{3} \begin{bmatrix} p^{\rm {H}}_1 & p^{\rm {H}}_1 & p^{\rm {H}}_1
  & p^{\rm {H}}_2
  & \dots & p^{\rm {H}}_{\NstatesHagg} & p^{\rm {H}}_{\NstatesHagg}
  \end{bmatrix}~.
\label{eq:PiExpansion}
\end{align}

The reader should check that applying the rote expansion method given here
results in the same HMM for the GM Process as we found in \S\ref{GraphicalExpansion}.  

\section{Correlation Functions from HMMs}
\label{sec:CFsFromHMMs}

At this point, with the process expressed as an \MachineABC, we can derive expressions for the CFs.

We introduce the family of cyclic-relation functions ${\xoperator} (\symbolABC)
\in \{\coperator(\symbolABC), \aoperator(\symbolABC),
\soperator(\symbolABC)\}$, where, for example:
\begin{equation}
  \coperator(\symbolABC) = \left\{ 
  \begin{array}{l l}
    B & \quad \textrm{if  } \, \symbolABC = A\\
    C & \quad \textrm{if  } \, \symbolABC = B\\
    A & \quad \textrm{if  } \, \symbolABC = C\\
  \end{array} \right..
\end{equation}
Thus, $\coperator(\symbolABC)$ is the cyclic permutation function.
Complementarily, $\aoperator(\symbolABC)$ performs anticyclic permutation among
$\symbolABC \in \{ A, B, C \}$; $\soperator(\symbolABC)$ performs the identity
operation among $\symbolABC \in \{ A, B, C \}$ and is suggestively denoted with
an `s' for \emph{sameness}. In terms of the absolute position of the MLs---\ie,
$\AlphabetABC = \{A, B, C\}$---the CFs directly relate to the products of
particular sequences of TMs. This perspective suggests a way to uncover the
precise relation between the CFs and the TMs. Using this, we then give a
closed-form expression for $\Qxin$ for any given HMM.

\subsection{CFs from TMs}
\label{subsec:CFsFromTMs}

To begin, let us first consider the meaning of $\QcThree$. In words, this is
the probability that two MLs separated by two intervening MLs are cyclically
related. Mathematically, we might start by writing this as:
\begin{align}
   \QcThree = \Pr(A**B) +  \Pr(B**C) + \Pr(C**A),
\end{align}
where $*$ is a wildcard symbol denoting an indifference for the symbol observed
in its place \footnote{While it is tempting to add the stipulation that no two
consecutive symbols can be the same, this will fall out naturally from $\QsOne
= 0$ via the transition-constraints built into the $ABC$-machine construction.}.
That is, $*$s denote marginalizing over the intervening MLs such that, for
example:
\begin{align}
       \Pr(A**B) = \sum_{\symbolABC_1 \in \AlphabetABC} 
                   \sum_{\symbolABC_2 \in \AlphabetABC} \Pr(A \symbolABC_{1} \symbolABC_{2} B).
\end{align}
Making use of the TM-formalism discussed previously, this becomes:
\begin{align}
   \Pr(A**B) &= \sum_{\symbolABC_{1} \in \AlphabetABC} \sum_{\symbolABC_{2} \in \AlphabetABC} 
               \Pr(A \symbolABC_{1} \symbolABC_{2} B) \nonumber \\
             &= \sum_{\symbolABC_{1} \in \AlphabetABC} \sum_{\symbolABC_{2} \in \AlphabetABC} 
               \bra \Dist  \TransMatABC^{[A]} \TransMatABC^{[\symbolABC_{1}]} 
               \TransMatABC^{[\symbolABC_{2}]} \TransMatABC^{[B]}  \DistOne \nonumber \\
             &= \bra \Dist  \TransMatABC^{[A]} \Bigl( \sum_{\symbolABC_{1} \in \AlphabetABC} 
                \sum_{\symbolABC_{2} \in \AlphabetABC} \TransMatABC^{[\symbolABC_{1}]} 
                \TransMatABC^{[\symbolABC_{2}]} \Bigr) \ \TransMatABC^{[B]}  \DistOne \nonumber \\
             &= \bra \Dist  \TransMatABC^{[A]} \Bigl( \underbrace{\sum_{\symbolABC_{1} \in \AlphabetABC} 
                \TransMatABC^{[\symbolABC_{1}]} }_{= \TransMatABC} \Bigr) 
                \Bigl( \underbrace{\sum_{\symbolABC_{2} \in \AlphabetABC} 
                \TransMatABC^{[\symbolABC_{2}]} }_{= \TransMatABC} \Bigr) \TransMatABC^{[B]}  \DistOne \nonumber \\
             &= \bra \Dist  \TransMatABC^{[A]} (\TransMatABC) (\TransMatABC) \TransMatABC^{[B]} \nonumber  \DistOne \\
             &= \bra \Dist  \TransMatABC^{[A]} \TransMatABC^2 \TransMatABC^{[B]}  \DistOne~, \nonumber 
\end{align}
where $\DistOne$ is a column vector of 1s of length $\NstatesABC$.
Hence, we can rewrite $\QcThree$ as:
\begin{align}
    \QcThree &= \Pr(A**B) +  \Pr(B**C) + \Pr(C**A) \nonumber \\
             &= \bra \Dist  \TransMatABC^{[A]} \TransMatABC^2 \TransMatABC^{[B]}  \DistOne + \nonumber
                \bra \Dist  \TransMatABC^{[B]} \TransMatABC^2 \TransMatABC^{[C]}  \DistOne  \\ 
             &   \qquad + \bra \Dist  \TransMatABC^{[C]} \TransMatABC^2 \TransMatABC^{[A]}  \DistOne \nonumber \\
             &= \sum_{\symbolABC \in \AlphabetABC} 
                \bra \Dist  \TransMatABC^{[\symbolABC]} \TransMatABC^2 
                \TransMatABC^{[\coperator(\symbolABC)]}  \DistOne. \nonumber 
\end{align}
For \emph{mixing} {\MachineABC}s, 
$\Pr(A**B) =  \Pr(B**C) = \Pr(C**A) = \frac{1}{3} \QcThree$, 
in which case the above reduces to:
\begin{align}
     \QcThree = 3 \bra \Dist  \TransMatABC^{[\symbolABC_{0}]} \TransMatABC^2 
               \TransMatABC^{[\coperator(\symbolABC_{0})]}  \DistOne, \text{ where } 
               \symbolABC_{0} \in \AlphabetABC. \nonumber 
\end{align}

The generalization to express any $\Qxin$ in terms of TMs may already be
obvious by analogy. Nevertheless, we give a brief derivation for completeness,
using similar concepts to those developed more explicitly above. For all $\xi
\in \{\rm{c}, \rm{a}, \rm{s} \}$ and for all $n \in \{1, 2, 3, \dots \}$, we can write the CFs
as:
\begin{align}
    \Qxin &= \Pr(A \underbrace{* \dots *}_{n-1 \text{ } * \text{s}} \xoperator (A)) 
                   + \Pr(B \underbrace{* \dots *}_{n-1 \text{ } * \text{s}} \xoperator (B)) \nonumber \\
                  &    \qquad  + \Pr(C \underbrace{* \dots *}_{n-1 \text{ } * \text{s}} \xoperator (C)) \nonumber \\
               &=  \sum_{\symbolABC_{0} \in \AlphabetABC} \Pr(\symbolABC_{0} 
                  \underbrace{* \dots *}_{n-1 \text{ } * \text{s}} \xoperator (\symbolABC_{0})) \nonumber \\
               &=  \sum_{\symbolABC_{0} \in \AlphabetABC} \sum_{\w \in \AlphabetABC{^{n-1}}}   
                   \Pr(\symbolABC_{0} \, \w \, \xoperator (\symbolABC_{0})) \nonumber \\
               &=  \sum_{\symbolABC_{0} \in \AlphabetABC} \sum_{\w \in \AlphabetABC^{n-1}} 
                   \bra \Dist  \TransMatABC^{[\symbolABC_{0}]} \TransMatABC^{[\w]} 
                   \TransMatABC^{[\xoperator (\symbolABC_{0})]} \DistOne \nonumber  \\
               &=  \sum_{\symbolABC_{0} \in \AlphabetABC} \bra \Dist 
                   \TransMatABC^{[\symbolABC_{0}]} \Bigl(\sum_{\w \in \AlphabetABC^{n-1}} 
                   \TransMatABC^{[\w]} \Bigr) \TransMatABC^{[\xoperator (\symbolABC_{0})]} \DistOne \nonumber \\
               &=  \sum_{\symbolABC_{0} \in \AlphabetABC} \bra \Dist  \TransMatABC^{[\symbolABC_{0}]} 
                  \Bigl(\prod_{i=1}^{n-1} 
                  \underbrace{\sum_{\symbolABC_{i} \in \AlphabetABC} 
                  \TransMatABC^{[\symbolABC_{i}]} }_{ = \TransMatABC} \Bigr) 
                  \TransMatABC^{[\xoperator (\symbolABC_{0})]} \DistOne \nonumber \\
              &=  \sum_{\symbolABC_{0} \in \AlphabetABC} \bra \Dist 
                 \TransMatABC^{[\symbolABC_{0}]} \TransMatABC^{n-1} 
                \TransMatABC^{[\xoperator (\symbolABC_{0})]}  \DistOne ,
  \label{Eq:TMsToCFs}
\end{align}
where the stationary distribution $\bra{\Dist}$ over states of the
$ABC$-machine is found from Eq.~(\ref{eq:ProbabilityDistribution}). 
\emph{The most general connection between CFs and TMs is given by Eq. (\ref{Eq:TMsToCFs}).} 

As before, we might assume on physical grounds that:
\begin{align}
 \label{eq:EqualProbs}
     \Pr(A \underbrace{* \dots *}_{n-1 \text{ } * \text{s}} \xoperator (A)) = 
     \Pr(B \underbrace{* \dots *}_{n-1 \text{ } * \text{s}} \xoperator (B)) = 
     \Pr(C \underbrace{* \dots *}_{n-1 \text{ } * \text{s}} \xoperator (C)).
\end{align}     
For example, Eq.~\eqref{eq:EqualProbs} is always true of mixing {\MachineABC}s. 
This special case yields the more constrained set of equations:
\begin{align}
\label{eq:ThreeTimesBraKetSimplification}
     \Qxin = 3 \bra \Dist  \TransMatABC^{[\symbolABC_{0}]} \TransMatABC^{n-1} 
             \TransMatABC^{[\xoperator (\symbolABC_{0})]}  \DistOne ~, 
\end{align}
where $\symbolABC_{0} \in \AlphabetABC$.

\subsection{CFs from Spectral Decomposition}
\label{subsec:SpectralResults}

Although Eq.~\eqref{Eq:TMsToCFs} is itself an important result, we can also
apply a spectral decomposition of powers of the TM to provide a closed-form
that is even more useful and insightful. Ameliorating the computational burden,
this result reduces the matrix powers in the above expressions to expressions
involving only powers of scalars. Also, yielding theoretical insight, the
closed-forms reveal what types of behaviors can ever be expected of the CFs
from stacking processes described by finite HMMs.

The most familiar case occurs when the TM is diagonalizable. Then,
$\TransMatABC^{n-1}$ can be found via diagonalizing the TM, making use of the
fact that $\abcT^L = C D^L C^{-1}$, given the eigen-decomposition $\abcT = C D
C^{-1}$, where $D$ is the diagonal matrix of eigenvalues. However, to
understand the CF behavior, it is more appropriate to decompose the matrix in
terms of its projection operators.

Moreover, an analytic expression for $\TransMatABC^{n-1}$ can be found in terms
of the projection operators even when the TM is not diagonalizable. Details are
given elsewhere~\cite{Crut13a,Riec14a}. By way of summarizing, though, in the
general case the $L^{\text{th}}$ iteration of the TM follows from:
\begin{align}
     \abcT^L = \mathcal{Z}^{-1} \left\{ \left( \IdentMat - z^{-1} \abcT \right) ^{-1} \right\}
     \label{eq:MatrixPowersViaZtransform}
  ~,
\end{align}
where $\IdentMat$ is the $\NstatesABC \times \NstatesABC$ identity matrix, $z
\in \mathbb{C}$ is a continuous complex variable, and $\mathcal{Z}^{-1} \{ \cdot \}$ denotes the inverse $z$-transform~\cite{Oppe75a} defined to operate elementwise: 
\begin{align}
\mathcal{Z}^{-1}\left( g(z) \right) & \equiv 
  \frac{1}{2 \pi i} \oint_{C} z^{L-1} g(z) \, dz 
\end{align}
for the $z$-dependent matrix element $g(z)$ of $\left(  \IdentMat - z^{-1}
\abcT \right)^{-1}$. Here, $\oint_C$ indicates a counterclockwise contour
integration in the complex plane enclosing the entire unit circle. 

For nonnegative integers $L$, and with the allowance that $0^L = \delta_{L,0}$
for the case that $0 \in \Lambda_{\abcT}$, Eq.\ \eqref{eq:MatrixPowersViaZtransform} becomes: 
\begin{align}
\abcT^L 
& = \sum_{\lambda \in \Lambda_{\abcT} } 
		\sum_{m=0}^{\nu_\lambda - 1}  
		\lambda^{L-m}  \binom{L}{m} 
		\abcT_\lambda  
		\left( \abcT - \lambda \IdentMat \right)^m
  ~,
\label{eq: T^L spectral decomp for positive integer L}		
\end{align}
where $\Lambda_{\abcT} = \{ \lambda \in \mathbb{C}: \text{det}(\lambda
\IdentMat - \abcT) = 0 \}$ is the set of $\abcT$'s eigenvalues, $\abcT_\lambda$
is the projection operator associated with the eigenvalue $\lambda$ given by
the elementwise residue of the resolvent $\left( z \IdentMat - \abcT
\right)^{-1}$ at $z \to \lambda$, the index $\nu_\lambda$ of the eigenvalue
$\lambda$ is the size of the largest Jordan block associated with $\lambda$,
and $\binom{L}{m} = \frac{L!}{m! (L-m)!}$ is the binomial
coefficient.\footnote{Recall, \eg, 
that $\binom{L}{0} = 1$, $\binom{L}{1} = L$, $\binom{L}{2} = \tfrac{1}{2!} L (L-1)$, 
and $\binom{L}{L} = 1$.}
In terms of elementwise contour integration, we have:
\begin{align}
\abcT_\lambda = \frac{1}{2 \pi i} \oint_{C_\lambda} \left( z \IdentMat - \abcT \right)^{-1} \, dz, 
\label{eq:GeneralProjOpEqn}
\end{align}
where $C_\lambda$ is any contour in the complex plane enclosing the point $z_0
= \lambda$---which may or may not be a singularity depending on the particular
element of the resolvent matrix---but encloses no other singularities.

As guaranteed by the Perron--Frobenius theorem, all eigenvalues of the
stochastic TM $\abcT$ lie on or within the unit circle. Moreover, the
eigenvalues on the unit circle are guaranteed to have index one. The indices
of all other eigenvalues must be less than or equal to one more than the
difference between their algebraic $a_\lambda$ and geometric $g_\lambda$
multiplicities. Specifically:
\begin{align*}
\nu_\lambda - 1 \leq a_\lambda - g_\lambda \leq a_\lambda - 1 ~\text{and}~
\nu_\lambda = 1 , \text{ if } |\lambda| = 1
  ~.
\end{align*}

Using Eq.~\eqref{eq: T^L spectral decomp for positive integer L} together with 
Eq.~\eqref{Eq:TMsToCFs}, the CFs can now be expressed as: 
\begin{align}
\Qxin & = \sum_{\lambda \in \Lambda_{\abcT}} 
  \sum_{m=0}^{\nu_\lambda - 1}  \CFBraKetm \binom{n-1}{m}  \lambda^{n-m-1} 
  ~,
\label{eq:GenCFdecomp}
\end{align}
where $\CFBraKetm$ is a complex-valued scalar:\footnote{$\CFBraKetm$ is
constant with respect to the relative layer displacement $n$. However, $\Big\{
\CFBraKetm \Big\}$ can be a function of a process's parameters.}
\begin{align} 
\CFBraKetm \equiv
\sum_{\symbolABC_{0} \in \AlphabetABC} \bra \Dist 
          \TransMatABC^{[\symbolABC_{0}]} 
          \abcT_\lambda  \left( \abcT - \lambda \IdentMat \right)^m
          \TransMatABC^{[\xoperator (\symbolABC_{0})]} \DistOne
		  ~.
\end{align}

Evidently, the CFs' mathematical form Eq.~\eqref{eq:GenCFdecomp} is strongly
constrained for any stacking process that can be described by a finite HMM.
Besides the expression's elegance, we note that its constrained form is very
useful for the so-called ``inverse problem'' of discovering the stacking
process from CFs~\cite{Varn02a,Varn07a,Varn13a,Varn13b}.

When $\abcT$ is diagonalizable, $\nu_\lambda = 1$ for all $\lambda$ so that 
Eq.~\eqref{eq: T^L spectral decomp for positive integer L} simply reduces to:
\begin{align}
\abcT^L 
& = \sum_{\lambda \in \Lambda_{\abcT} } 
	\lambda^L  \abcT_\lambda  
  ~,
\end{align}
where the projection operators can be obtained more simply as: 
\begin{align}
\label{eq: T_lambda algorithm}
\abcT_{\lambda} & = \prod_{\substack{\zeta \in \Lambda_{\abcT} \\ \zeta \neq \lambda }}
  \frac{\abcT - \zeta \IdentMat }{\lambda  - \zeta}
  ~.
\end{align}
In the diagonalizable case, Eq.~\eqref{eq:GenCFdecomp} reduces to: 
\begin{align}
\Qxin &=  \sum_{\lambda \in \Lambda_{\abcT}} \lambda^{n-1} 
  \sum_{\symbolABC_{0} \in \AlphabetABC} \bra \Dist 
         \TransMatABC^{[\symbolABC_{0}]} \abcT_\lambda
         \TransMatABC^{[\xoperator (\symbolABC_{0})]} \DistOne \nonumber \\
& = \sum_{\lambda \in \Lambda_{\abcT}} 
 \CFBraKet  \lambda^{n-1}
  ~,
\label{eq:DiagCFdecomp}
\end{align}
where 
$\CFBraKet \equiv \CFBraKeto$ is again a constant: 
\begin{align}
\CFBraKet & = \sum_{\symbolABC_{0} \in \AlphabetABC} \bra \Dist 
        \TransMatABC^{[\symbolABC_{0}]} \abcT_\lambda  
        \TransMatABC^{[\xoperator (\symbolABC_{0})]}  \DistOne
  ~.
\label{eq:DiagonalizableCorrelationAmplitudes}
\end{align}

\subsection{Asymptotic behavior of the CFs}

From the spectral decomposition, it is apparent that the CFs converge to
some constant value as $n \to \infty$, unless $\abcT$ has eigenvalues 
on the unit circle besides unity itself.
If unity is the sole eigenvalue with a magnitude of one, then all
other eigenvalues have magnitude less than unity and their contributions decay to
negligibility for large enough $n$.  Explicitly, if $\argmax_{\lambda \in
\Lambda_{\abcT}} |\lambda| = \{ 1 \}$, then:
\begin{align*}
\lim_{n \to \infty} \Qxin 
& = \lim_{n \to \infty} 
    \sum_{\lambda \in \Lambda_{\abcT}} 
    \sum_{m=0}^{\nu_\lambda - 1}  
       \CFBraKetm \binom{n-1}{m}  \lambda^{n-m-1} \\
& = \Braket{\abcT_1 ^{\xi(\Alphabet)}} \\ 
& = \sum_{\symbolABC_{0} \in \AlphabetABC} \bra \Dist 
                 \TransMatABC^{[\symbolABC_{0}]} 
                 \abcT_1  
                \TransMatABC^{[\xoperator (\symbolABC_{0})]}  \DistOne \\ 
& = \sum_{\symbolABC_{0} \in \AlphabetABC} \bra \Dist 
                 \TransMatABC^{[\symbolABC_{0}]} 
                 \DistOne  \bra \Dist
                \TransMatABC^{[\xoperator (\symbolABC_{0})]}  \DistOne \\
& = \sum_{\symbolABC_{0} \in \AlphabetABC}
	\Pr(\symbolABC_{0}) \Pr(\xoperator (\symbolABC_{0})) 
  ~.
\end{align*}
Above, we used the fact that $\nu_1 = 1$ and that---for an ergodic
process---$\abcT_1 = \DistOne \bra \Dist$.

For mixing $ABC$-machines, $\Pr(\symbolABC) = 1 / 3$
for all $\symbolABC \in \AlphabetABC$. That this is so should
be evident from the graphical expansion method of \S\ref{GraphicalExpansion}.
Therefore, mixing processes with $\argmax_{\lambda \in
\Lambda_{\abcT}} |\lambda| = \{ 1 \}$ have CFs that all converge to $1/3$:
\begin{align*}
\lim_{n \to \infty} \Qxin 
& = \sum_{\symbolABC_{0} \in \AlphabetABC}
	\Pr(\symbolABC_{0}) \Pr(\xoperator (\symbolABC_{0})) \\
& = 3 ( \tfrac{1}{3} \times \tfrac{1}{3} ) \\ 
& = \tfrac{1}{3}
  ~.
\end{align*}

Non-mixing processes with $\argmax_{\lambda \in
\Lambda_{\abcT}} |\lambda| = \{ 1 \}$ can have their CFs converging to
constants other than $1/3$, depending on $\{ \Pr(\symbolABC) : \symbolABC \in \AlphabetABC \}$,
although they are still constrained by $\sum_{\xi} \Qxin = 1$.

If other eigenvalues in $\Lambda_{\TransMatABC}$ beside unity exist on the unit circle, then the CFs approach a periodic sequence as $n$ gets large. 

\subsection{Modes of Decay}

Since $\TransMatABC$ has no more eigenvalues than its dimension (\ie,
$|\Lambda_{\TransMatABC}| \leq \NstatesABC$), {\emph{Eq.~\eqref{eq:GenCFdecomp}
implies that the number of states in the \MachineABC\ for a stacking process
puts an upper bound on the number of modes of decay.}}  Indeed, since unity is
associated with stationarity, the number of modes of decay is strictly less
than $\NstatesABC$.  It is important to note that these modes do not always
decay strictly exponentially: They are in general the product of a decaying
exponential with a polynomial in $n$, and the CFs are sums of these products. 

Even if---due to diagonalizability of $\TransMatABC$---there were only strictly
exponentially decaying modes, it is simple but important to understand that
there is generally more than one mode of exponential decay present in the CFs.
And so, ventures to find \emph{the} decay constant of a process are misleading
unless it is explicitly acknowledged that one seeks, \eg, the slowest decay
mode. Even then, however, there are cases when the slowest decay mode only acts
on a component of the CFs with negligible magnitude.  In an extreme case, the
slowest decay mode may not even be a large contributor to the CFs before the
whole pattern is numerically indistinguishable from the asymptotic value.

In analyzing a broad range of correlation functions, nevertheless, many authors have been led to consider \emph{correlation lengths}, also known as \emph{characteristic lengths}~\cite{Tiwa07a,Varn13b}. The form of Eq.~\eqref{eq:GenCFdecomp} suggests that this perspective will often be a clumsy oversimplification for understanding CFs. Regardless, if one wishes to assign a correlation length associated with an index-one mode of CF decay, we observe that the reciprocal of the correlation length is essentially the negative logarithm of the magnitude of the eigenvalue for that mode. We find that the typically reported correlation length $\ell_{\text{C}}$ derives from the
second-largest contributing magnitude among the eigenvalues:
\begin{align}
\ell_{\text{C}}^{-1} & = -\log |\zeta| ~,
  & \text{for } \zeta \in \argmax_{\lambda \in \boldsymbol{\rho} } |\lambda|~,
\label{eq:CorrLen}
\end{align}
where $\boldsymbol{\rho} = \left\{ \lambda \in \Lambda_{\abcT} \setminus \{ 1 \} : \CFBraKet \neq 0 \right\}$.
%Phase transitions correspond to 
%collisions and subsequent scattering of eigenvalues as parameters are tweaked, 
%with an ephemeral nondiagonalizability through that passage leading to infinite correlation length. 

Guided by Eq.~\eqref{eq:GenCFdecomp}, we suggest that a true understanding of CF behavior 
involves finding $\Lambda_{\TransMatABC}$ with the corresponding eigenvalue indices and the amplitude of each mode's contribution $\Big\{ \CFBraKetm \Big\}$. 

This now completes our theoretical development, and in the next section we apply these
techniques to three examples.

\section{Examples}
\label{Examples}

\subsection{3C Polytypes and Random ML Stacking: IID Processes}
\label{RandomStacking}

Although not often applicable in practice, as a pedagogical exercise the random
ML stacking has often been treated~\cite{Guin63a}. This stacking process is the
simplest stacking arrangement that can be imagined,\footnote{This is not mere
hyperbole. It is possible to quantify a process's structural organization in
the form of its \emph{statistical complexity} $\Cmu$, which measures the
internal information processing required to produce the
pattern~\cite{Crut89a,Crut12a,Varn13a}. In the present case $\Cmu = 0$ bits,
the minimum value.} and there are previous analytical results that can be
compared to the techniques developed here. In statistics parlance, this process
is an \emph{independent and identically distributed} (IID)
process~\cite{Cove06a}.

\begin{figure}
\begin{center}
\includegraphics[width=0.25\textwidth]{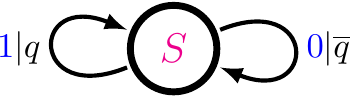}
\end{center}
\caption{\MachineHagg\ for the IID Process. When $q=1$, the IID Process
  generates a string of $1$s, which is physically the 3C$^+$ stacking
  structure. Conversely, when $q=0$, the structure corresponds to the 3C$^-$
  structure. For $q = 1/2$, the MLs are stacked as randomly as possible.
  }
\label{FigRandomProcessHagg}
\end{figure}

Let us assume that the placement of MLs is independent of the previous MLs
scanned, except that it of course must obey the stacking constraints. The
\MachineHagg\ that describes this process is shown in
Fig.~\ref{FigRandomProcessHagg}. We allow for the possibility that there might
be a bias in the stacking order, and we assign a probability $q$ that the next
layer is cyclically related to its predecessor. Thus, the 1-by-1 symbol-labeled
TMs for the \MachineHagg\ are:
\begin{align*}
  \begin{array}{r@{\mskip\thickmuskip}l}
     \HaggTone & = \begin{bmatrix} q \end{bmatrix}   
  \end{array} 
  {\rm {and}}
  \begin{array}{r@{\mskip\thickmuskip}l}
       \HaggTzero = \begin{bmatrix} \overline{q} \end{bmatrix} ,
   \end{array}
\end{align*}
where $\overline{q} \equiv 1-q$, with $q \in [0,1]$.

The physical interpretation of the IID Process is straightforward. In the case
where $q = 1$, the process generates a stacking sequence of all $1$s, giving a
physical stacking structure of $\dots ABCABCABC \dots$. We recognize this as
the 3C$^{+}$ crystal structure. Similarly, for $q = 0$, the process generates
stacking sequence of all 0s, which is the 3C$^{-}$ crystal structure. For those
cases where $q$ is near but not quite at its extreme values, the stacking
structure is 3C with randomly distributed deformation faults. When $q=\OneHalf$, the
MLs are stacked in a completely random fashion.

\begin{figure}
\begin{center}
\includegraphics[width=0.45\textwidth]{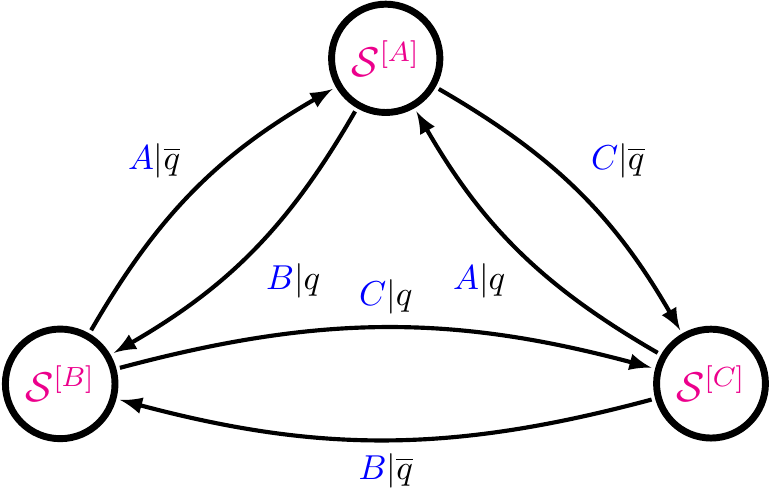}
\end{center}
\caption{\MachineABC\ for the IID Process. The single state of the
  \MachineHagg\ has expanded into three.
  }
\label{FigRandomProcessABC}
\end{figure}

Now, we must determine whether this is a mixing or nonmixing \MachineHagg. We
note that there are two SSCs, namely [$\StateHagg_{(0)}$] and
[$\StateHagg_{(1)}$]. The winding numbers for each are
$W^{[\StateHagg_{(1)}]} = 1$ and $W^{[\StateHagg_{(0)}]} = 2$, respectively.
Since at least one of these is not equal to zero, the \MachineHagg\ is mixing,
and we need to expand the \MachineHagg\ into the \MachineABC. This is shown in
Fig.~\ref{FigRandomProcessABC}.

The \MachineABC\ TMs can either be directly written down from inspecting
Fig.~\ref{FigRandomProcessABC} or by using the rote expansion algorithm of
\S\ref{AlgebraicExpansion}, using Eqs.~\eqref{eq:T0toABC}
and~\eqref{eq:T1toABC}. By either method we find the 3-by-3 TMs to be:
\begin{align*}
  \begin{array}{r@{\mskip\thickmuskip}l}
\abcTA
=
\begin{bmatrix}
0 & 0 & 0 \\
\overline{q} & 0 & 0 \\
q & 0 & 0
\end{bmatrix} ,
  \end{array} 
  \begin{array}{r@{\mskip\thickmuskip}l}
\abcTB
=
\begin{bmatrix}
0 & q & 0 \\
0 & 0 & 0 \\
0 & \overline{q} & 0
\end{bmatrix}
          \end{array}
\end{align*}
and
\begin{align}
\abcTC
=
\begin{bmatrix} \nonumber
0 & 0 & \overline{q} \\
0 & 0 & q \\
0 & 0 & 0
\end{bmatrix} .
\end{align}
The internal state TM then is their sum:
\begin{align}
\abcT
=
\begin{bmatrix}  \nonumber
0 & q & \overline{q} \\
\overline{q} & 0 & q \\
q & \overline{q} & 0
\end{bmatrix}
  ~.
\end{align}
The eigenvalues of the $ABC$ TM are 
\begin{align*}
\Lambda_{\abcT} = \{ 1, \, \Omega, \, \Omega^* \}~,
\end{align*}
where:
\begin{align*}
\Omega \equiv - \frac{1}{2} + i \frac{\sqrt{3}}{2} (4q^2 - 4q + 1 ) ^{1/2}
\end{align*}
and $\Omega^*$ is its complex conjugate. 
Already, via Eq.~\eqref{eq:CorrLen}, we can identify what the characteristic length of the CFs will be.
In particular, $\ell_{\text{C}}^{-1} = -\log |\Omega| = -\tfrac{1}{2} \log (1 - 3q + 3q^2)$ yields:
\begin{align*}
\ell_{\text{C}} = -\frac{2}{\log(1 - 3q + 3q^2)}~.
\end{align*}
If we identify $q$ with the deformation faulting parameter $\alpha$ in the model introduced
by Estevez \etal~\cite{Este08a} (see the next example in \S\ref{GrowthDeformationStacking}, the RGDF Process), 
this is identical to the result obtained there
in Eq. (35).  There is much more structural information in the CFs, however,
than a single characteristic length would suggest.  This fact will become 
especially apparent as our examples become more sophisticated. 

According to Eq.~\eqref{Eq:TMsToCFs}, we can obtain the CFs via:
\begin{align}
\Qxin &=  \sum_{\symbolABC_{0} \in \AlphabetABC} \bra \Dist 
                 \TransMatABC^{[\symbolABC_{0}]} \TransMatABC^{n-1} 
                \TransMatABC^{[\xoperator (\symbolABC_{0})]}  \DistOne . \nonumber
\end{align}
The stationary distribution over the $ABC$-machine states is found from 
Eq.~\eqref{eq:ProbabilityDistribution}:
\begin{align*}
\bra{\Dist}  = \left[ \tfrac{1}{3} \quad \tfrac{1}{3}  \quad \tfrac{1}{3} \right]
  ~.
\end{align*}
Furthermore, an analytic expression for $\abcT^{n-1}$ follows
from the $z$-transform as given in Eq.~\eqref{eq:MatrixPowersViaZtransform}.
As a start, we find:
\begin{align*}
\IdentMat - z^{-1} \abcT =
\begin{bmatrix}
1 & -q/z & -\overline{q}/z \\
-\overline{q}/z & 1 & -q/z \\
-q/z & -\overline{q}/z & 1
\end{bmatrix} 
\end{align*}
and its inverse:
\begin{align*}
\left( \IdentMat - z^{-1} \abcT \right) ^{-1} & =   
\frac{1}{(1-z^{-1})(1- \Omega z^{-1})(1- \Omega ^*z^{-1})} \\ 
  \times & \begin{bmatrix}
1- q \overline{q} z^{-2} 			& qz^{-1}+ \overline{q}^2 z^{-2} 		& \overline{q} z^{-1}+ q^2 z^{-2} \\
\overline{q} z^{-1}+ q^2 z^{-2}	& 1- q \overline{q} z^{-2} 			& qz^{-1}+ \overline{q}^2 z^{-2} \\
qz^{-1}+ \overline{q}^2 z^{-2} 	& \overline{q} z^{-1}+ q^2 z^{-2} 		& 1- q \overline{q} z^{-2}
\end{bmatrix} .
\end{align*}
%where:
%\begin{align*}
%\Omega \equiv - \frac{1}{2} + i \frac{\sqrt{3}}{2} (4q^2 - 4q + 1 ) ^{1/2}
%\end{align*}
%and $\Omega^*$ is its complex conjugate\footnote{For comparison with the 
%spectral perspective of \S \ref{subsec:SpectralResults}, note that the eigenvalues of the $ABC$ TM are $\Lambda_{\abcT} = \{ 1, \, \Omega, \, \Omega^* \}$.}. 
Upon partial fraction expansion, we obtain:
\begin{align}
& \left( \IdentMat - z^{-1} \abcT \right) ^{-1} \nonumber \\
  & =   \frac{1}{3} \frac{1}{(1-z^{-1})} 
\begin{bmatrix}
1	& 1 		& 1 \\
1	& 1 		& 1 \\
1	& 1 		& 1 
\end{bmatrix} \nonumber \\ 
& +
\tfrac{1}{(\Omega -1) (\Omega - \Omega ^*)}
\tfrac{1}{(1- \Omega z^{-1})} 
\begin{bmatrix}
\Omega ^2 - q \overline{q} 		& q \Omega + \overline{q}^2 		& \overline{q} \Omega + q^2 \\
\overline{q} \Omega+ q^2		& \Omega ^2 - q \overline{q}		& q \Omega + \overline{q}^2 \\
q \Omega + \overline{q}^2 	 	& \overline{q} \Omega + q^2  		& \Omega ^2 - q \overline{q}
\end{bmatrix} \nonumber \\ 
& +
\tfrac{1}{(\Omega^* -1) (\Omega^* - \Omega)}
\tfrac{1}{(1- \Omega^* z^{-1})} 
\begin{bmatrix}
{\Omega^*} ^2 - q \overline{q} 	& q \Omega^* + \overline{q}^2 		& \overline{q} \Omega^* + q^2 \\
\overline{q} \Omega^*+ q^2	& {\Omega^*} ^2 - q \overline{q}		& q \Omega^* + \overline{q}^2 \\
q \Omega^* + \overline{q}^2 	& \overline{q} \Omega^* + q^2  		& {\Omega^*} ^2 - q \overline{q}
\end{bmatrix} 
  ,
\label{eq:RandomResolvent}
\end{align}
for $q \neq 1/2$. (The special case of $q = 1/2$ is discussed in the next
subsection.) Finally, we take the inverse $z$-transform of
Eq.~\eqref{eq:RandomResolvent} to obtain an expression for the $L^{\text{th}}$
iterate of the TM:
\begin{align*}
\abcT^L & = \mathcal{Z}^{-1} 
  \left\{ \left( \IdentMat - z^{-1} \abcT \right) ^{-1} \right\} \\
  & = 
\frac{1}{3}
\begin{bmatrix}
1	& 1 		& 1 \\
1	& 1 		& 1 \\
1	& 1 		& 1 
\end{bmatrix} \\ 
& ~~+
\frac{\Omega^L}{(\Omega -1) (\Omega - \Omega ^*)}
\begin{bmatrix}
\Omega ^2 - q \overline{q} 		& q \Omega + \overline{q}^2 		& \overline{q} \Omega + q^2 \\
\overline{q} \Omega+ q^2		& \Omega ^2 - q \overline{q}		& q \Omega + \overline{q}^2 \\
q \Omega + \overline{q}^2 	 	& \overline{q} \Omega + q^2  		& \Omega ^2 - q \overline{q}
\end{bmatrix} \\ 
& ~~+
\frac{{\Omega^*}^L}{(\Omega^* -1) (\Omega^* - \Omega)}
\begin{bmatrix}
{\Omega^*} ^2 - q \overline{q}	& q \Omega^* + \overline{q}^2 		& \overline{q} \Omega^* + q^2 \\
\overline{q} \Omega^*+ q^2	& {\Omega^*} ^2 - q \overline{q}		& q \Omega^* + \overline{q}^2 \\
q \Omega^* + \overline{q}^2 	& \overline{q} \Omega^* + q^2  		& {\Omega^*} ^2 - q \overline{q}
\end{bmatrix}
  .
\end{align*}

These pieces are all we need to calculate the CFs. Let's start with $\Qsn$.
First, we find:
\begin{align}
\bra \Dist  \abcT^{[A]} =
\begin{bmatrix}     \nonumber
\frac{1}{3} & 0 & 0
\end{bmatrix}
\end{align}
and:
\begin{align}
\abcT^{[\soperator(A)]} \DistOne = \abcT^{[A]} \DistOne = 
\begin{bmatrix}  \nonumber
0 \\
\overline{q} \\
q 
\end{bmatrix} .
\end{align}
Then:
\begin{align}
\bra \Dist & \abcT^{[A]} \abcT^{n-1} = \tfrac{1}{9}
\begin{bmatrix}
1 & 1 & 1
\end{bmatrix} \nonumber \\
+ &
\tfrac{1}{3}
\tfrac{\Omega^{n-1}}{(\Omega -1) (\Omega - \Omega ^*)}
\begin{bmatrix}
\Omega ^2 - q \overline{q} 		& q \Omega + \overline{q}^2 		& \overline{q} \Omega + q^2
\end{bmatrix} \nonumber \\
+ &
\tfrac{1}{3}
\tfrac{{\Omega^*}^{n-1}}{(\Omega^* -1) (\Omega^* - \Omega)}
\begin{bmatrix}
{\Omega^*} ^2 - q \overline{q} 	& q \Omega^* + \overline{q}^2 		& \overline{q} \Omega^* + q^2
\end{bmatrix}
\label{eq:RandomPiATnBra}
\end{align}
and:
\begin{align}
\bra \Dist \abcT^{[A]} \abcT^{n-1}  \abcT^{[A]}  \DistOne & =
\tfrac{1}{9} 
+ \tfrac{1}{3}
\tfrac{\Omega^{n-1}}{(\Omega -1) (\Omega - \Omega ^*)}
\left( 2 q \overline{q} \Omega + \Omega \Omega^* \right) \nonumber \\
+ &
\tfrac{1}{3}
\tfrac{{\Omega^*}^{n-1}}{(\Omega^* -1) (\Omega^* - \Omega)}
\left( 2 q \overline{q} \Omega^* + \Omega \Omega^* \right) .
\end{align}
One can verify that Eq.~\eqref{eq:ThreeTimesBraKetSimplification} can be applied in lieu of 
Eq.~\eqref{Eq:TMsToCFs}, which saves some effort in finding the final result,
which is:
\begin{align}
\Qsn = 1/3 & + 
2 \text{Re} \left\{ {
\frac{\Omega^{n}}{(\Omega -1) (\Omega - \Omega^*)}
\left( 2 q \overline{q} + \Omega^* \right) 
} \right\}
  .
\label{eq:generalIIDQs}
\end{align}

The cyclic and anticyclic CFs can also be calculated from
Eq.~\eqref{eq:ThreeTimesBraKetSimplification} using the result we have already
obtained in Eq.~\eqref{eq:RandomPiATnBra} and a quick calculation yields:
\begin{align}
\abcT^{[\coperator(A)]} \DistOne = \abcT^{[B]} \DistOne = 
\begin{bmatrix}   \nonumber
q \\
0 \\
\overline{q} \\
\end{bmatrix} 
\end{align}
and
\begin{align}
\abcT^{[\aoperator(A)]} \DistOne = \abcT^{[C]} \DistOne = 
\begin{bmatrix}  \nonumber
\overline{q} \\
q \\
0
\end{bmatrix}  .
\end{align}
Then, we have:
\begin{align}
\Qcn &= 3 \bra \Dist \abcT^{[A]} \abcT^{n-1} \abcT^{[B]} \DistOne \nonumber \\
&= 
1/3 
+ 
2 \text{Re} \left\{ {
\frac{\Omega^{n}}{(\Omega -1) (\Omega - \Omega^*)}
\left( \overline{q}^2 + q \Omega \right) 
} \right\} 
\label{eq:RandomStackingQcn}
\end{align}
and:
\begin{align}
\Qan &= 3 \bra \Dist  \abcT^{[A]} \abcT^{n-1} \abcT^{[C]} \DistOne \nonumber \\
&= 
1/3 
+ 
2 \text{Re} \left\{ {
\frac{\Omega^{n}}{(\Omega -1) (\Omega - \Omega^*)}
\left( q^2 + \overline{q} \Omega \right) 
} \right\} .
\label{eq:RandomStackingQan}
\end{align}

All of this subsection's results hold for the whole range of $q \in [0,
\tfrac{1}{2}) \cup (\tfrac{1}{2}, 1]$, where all $\abcT$'s eigenvalues are
distinct. However, for $q = 1 / 2$, the two complex conjugate
eigenvalues, $\Omega$ and $\Omega^*$, lose their imaginary components, becoming
repeated eigenvalues. This requires special treatment.\footnote{Indeed, the
straightforward $z$-transform approach yielding the CF equations given in this
section appears to need special treatment for $q = 1/2$. However, a
more direct spectral perspective as developed in \S
\ref{subsec:SpectralResults} shows that since $\abcT$ is diagonalizable for all
$q$, all eigenvalues have index of one and so yield CFs of the simple form of
Eq.~\eqref{eq:DiagCFdecomp}.} We address the case of $q = 1 / 2$ in the
next subsection, which is of interest in its own right as being the most random
possible stacking sequence allowed.

\subsubsection{A Fair Coin?}

When a close-packed structure has absolutely no underlying crystal order in
the direction normal to stacking, the stacking sequence is as random as it
possibly can be. This is the case of $q = 1/2$, where spins are effectively
assigned by a fair coin, which yields a symmetric TM with repeated eigenvalues.
Due to repeated eigenvalues, the CFs at least superficially obtain a special
form.

To obtain the CFs for the Fair Coin IID Process, we follow the procedure of the previous subsection, with all of the same results through Eq.~\eqref{eq:RandomResolvent}, which with $q = 1/2$ and $\left. \Omega \right|_{q = 1/2} = \left. \Omega^* \right|_{q = 1/2} = -1/2$ can now be written as:
\begin{align}
\left( \IdentMat - z^{-1} \abcT \right) ^{-1} 
& =  
\frac{1}{(1-z^{-1})(1+ \frac{1}{2} z^{-1})^2}  \nonumber \\ 
\times
&
\begin{bmatrix}  \nonumber
1-  \frac{1}{4} z^{-2} 	&  \frac{1}{2} z^{-1} +  \frac{1}{4} z^{-2} &  \frac{1}{2} z^{-1} +  \frac{1}{4} z^{-2} \\
 \frac{1}{2} z^{-1} +  \frac{1}{4} z^{-2}	& 1-  \frac{1}{4} z^{-2}   &  \frac{1}{2} z^{-1}+  \frac{1}{4} z^{-2} \\
 \frac{1}{2} z^{-1}+  \frac{1}{4} z^{-2} 	&  \frac{1}{2} z^{-1}+  \frac{1}{4} z^{-2}   & 1-  \frac{1}{4} z^{-2}
\end{bmatrix} .
\end{align}
However, the repeated factor in the denominator yields a new partial fraction
expansion. Applying the inverse $z$-transform gives the $L^{\text{th}}$ iterate
of the TM\footnote{By inspection, we see from Eq.~\eqref{eq:DiagCFdecomp} that
$\abcT^0$ is the identity matrix and $\abcT^1 = \abcT$, as must be the case.
More interestingly, the decaying deviation from the asymptotic matrix is
oscillatory.} as:
\begin{align*}
\abcT^L = \mathcal{Z}^{-1} 
&
\left\{ \left( \IdentMat - z^{-1} \abcT \right) ^{-1} \right\} \\
= \frac{1}{3}
&
\begin{bmatrix}  \nonumber
1	& 1 		& 1 \\
1	& 1 		& 1 \\
1	& 1 		& 1 
\end{bmatrix}
+ \frac{1}{3}
\left(- \frac{1}{2}\right)^L
\begin{bmatrix}
2 	& -1 		& -1 \\
-1	& 2		& -1 \\
-1	& -1  		& 2
\end{bmatrix}
  ~.
\end{align*}
Then, we find:
\begin{align}
\bra \Dist \abcT^{[A]} \abcT^{n-1} =
\frac{1}{9}
\begin{bmatrix} \nonumber
1 & 1 & 1
\end{bmatrix} 
+
\frac{1}{9}
\left(- \frac{1}{2}\right)^{n-1}
\begin{bmatrix}
2 & -1 & -1
\end{bmatrix} 
  .
\end{align}
With the final result that:
\begin{align}
\Qsn &= 3 \bra \Dist \abcT^{[A]} \abcT^{n-1} \abcT^{[A]} \DistOne \nonumber \\
&= 
\frac{1}{3} 
+ 
\frac{2}{3}  \left(- \frac{1}{2}\right)^{n} ,
\label{eq:IIDQs}
\end{align}
\begin{align}
\Qcn &= 3 \bra \Dist \abcT^{[A]} \abcT^{n-1} \abcT^{[B]} \DistOne \nonumber \\
&= 
\frac{1}{3} 
- 
\frac{1}{3}  \left(- \frac{1}{2}\right)^{n} ,
\end{align}
and
\begin{align}
\Qan &= 3 \bra \Dist \abcT^{[A]} \abcT^{n-1} \abcT^{[C]} \DistOne \nonumber \\
&= 
\frac{1}{3} 
- 
\frac{1}{3}  \left(- \frac{1}{2}\right)^{n} .
\end{align}
For $q = 1/2$, we see that $\Qcn$ and $\Qan$ are identical, but this is not
generally the case as one can check for other values of $q$ in
Eqs.~\eqref{eq:RandomStackingQcn} and~\eqref{eq:RandomStackingQan}.

\begin{figure}
\begin{center}
\includegraphics[width=0.5\textwidth]{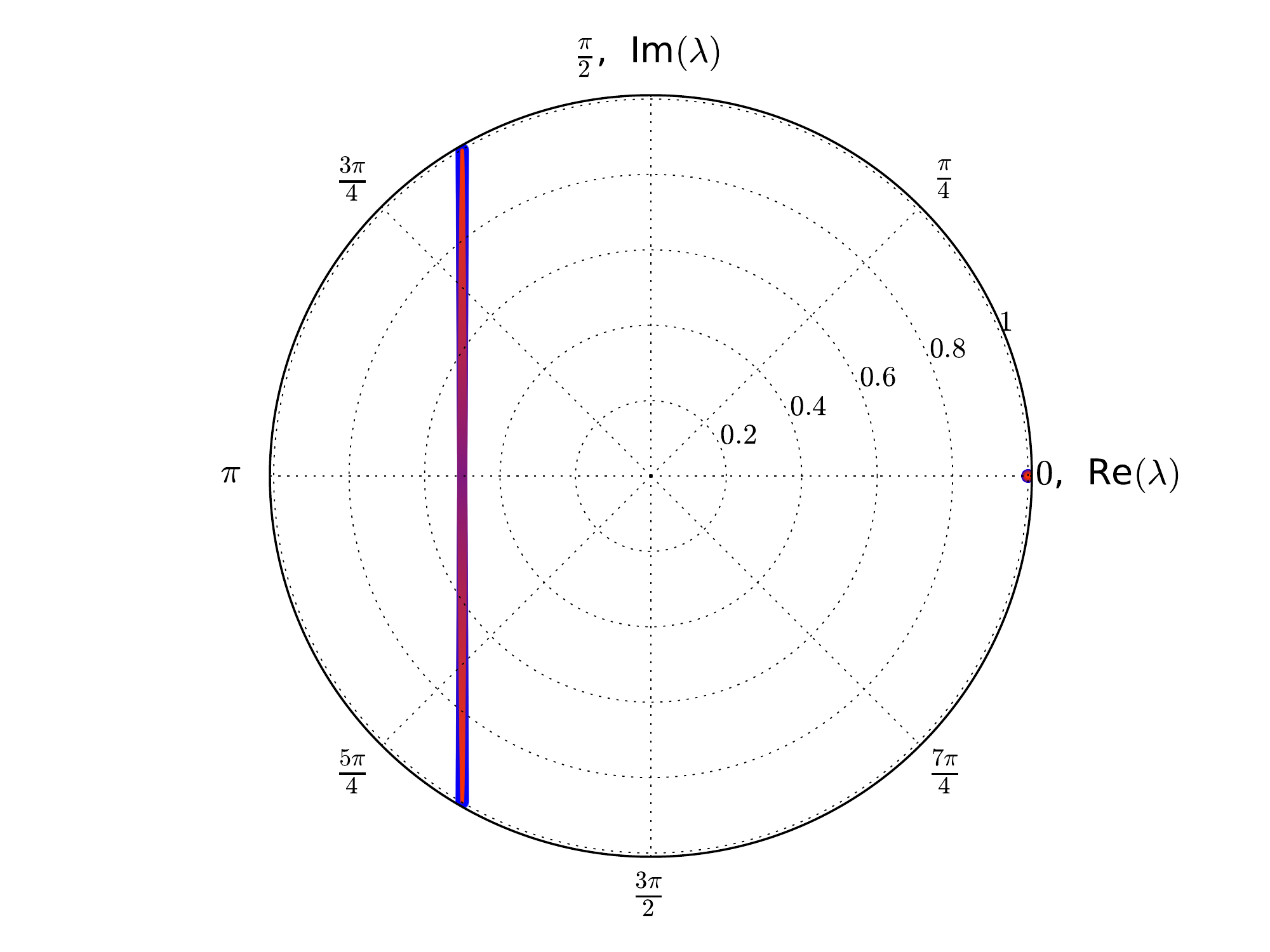}
\end{center}
\caption{TM's eigenvalues in the complex plane for the IID Process as $q$
  is varied. Note that there is always an eigenvalue at $1$.
  }
\label{fig:BernoulliABCTeigs}
\end{figure}

Figure~\ref{fig:BernoulliABCTeigs} shows a graph of the TM's eigenvalues in the
complex plane as $q$ is varied. Notice that there is an eigenvalue at $1$ for
all values of $q$. This is generic feature, and we always find such an
eigenvalue. The other two eigenvalues start at the other two cube roots of
unity for $q \in \{0,1\}$ and, as $q \to 1/2$, they migrate to the point $-1/2$
and become degenerate at $q = 1/2$. It is this degeneracy that requires the
special treatment given in this section.

It is interesting that even the Fair Coin \MachineHagg\ produces structured
CFs. This is because---even though the allowed transitions of the underlying
$ABC$-machine are randomized---not all transitions are allowed. For example, if
we start with an $A$ ML, the next ML has a zero probability of being an $A$, a
$1/2$ probability of being a $B$, and a $1/2$ probability of being a $C$. Then,
the \emph{next} ML has a rebounding $1/2$ probability of being an $A$ while the
probability of being either a $B$ or $C$ is each only $1/4$. So, we see that the
underlying process has structure, and there is nothing we can do---given the
physical constraints---to make the CFs completely random.

When we can compare our expressions for CFs at $q = 1/2$ to those derived
previously by elementary means~\cite{Guin63a,Varn01b}, we find agreement. Note
however that unlike in these earlier treatments, here there was no need to
assume a recursion relationship.

\begin{figure}
\begin{center}
\includegraphics[width=0.5\textwidth]{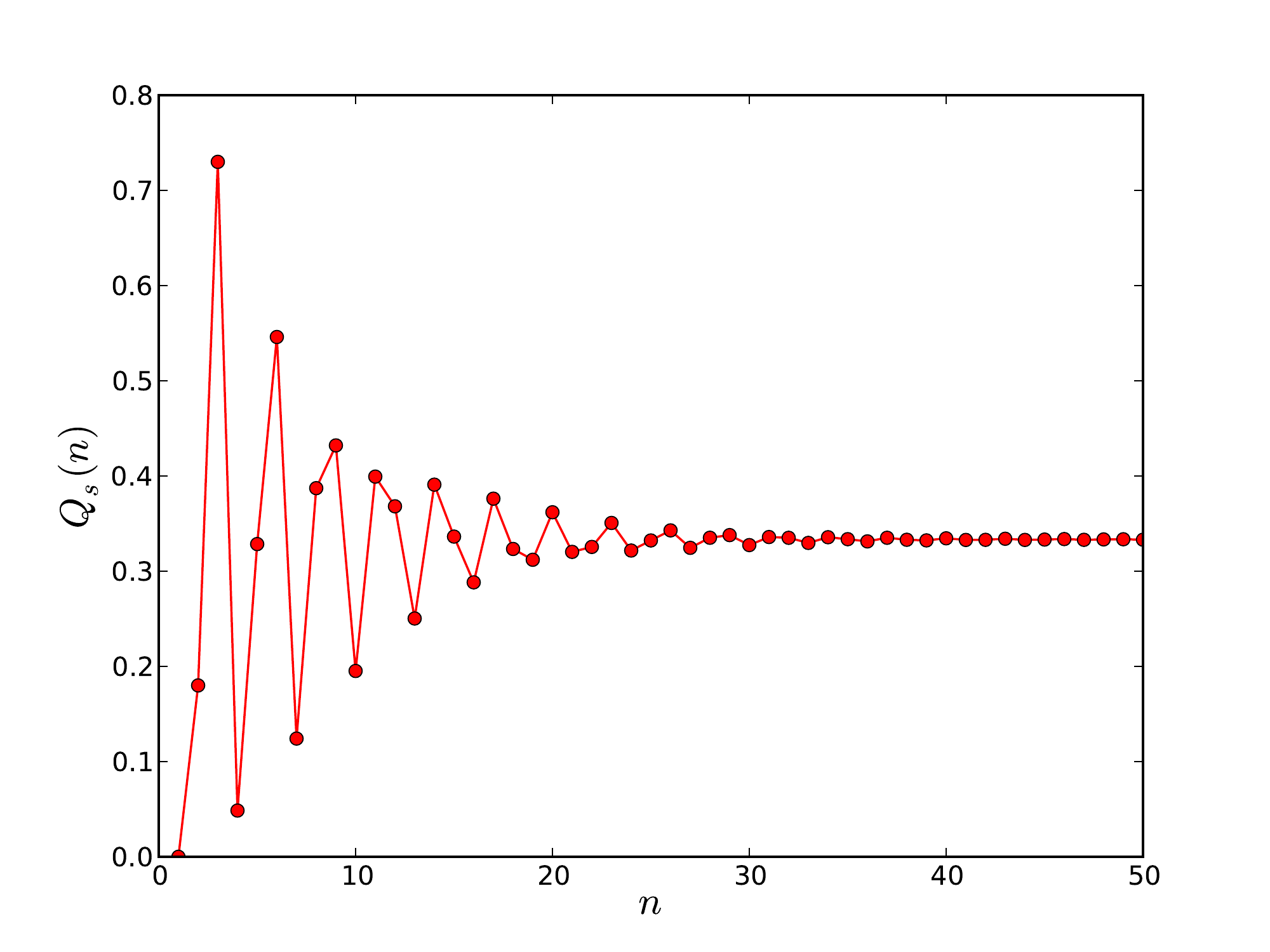}
\end{center}
\caption{$\Qsn$ vs. $n$ for $q=0.1$ the IID Process.}
\label{fig:BernoulliQs0p1q}
\end{figure}

\begin{figure}
\begin{center}
\includegraphics[width=0.5\textwidth]{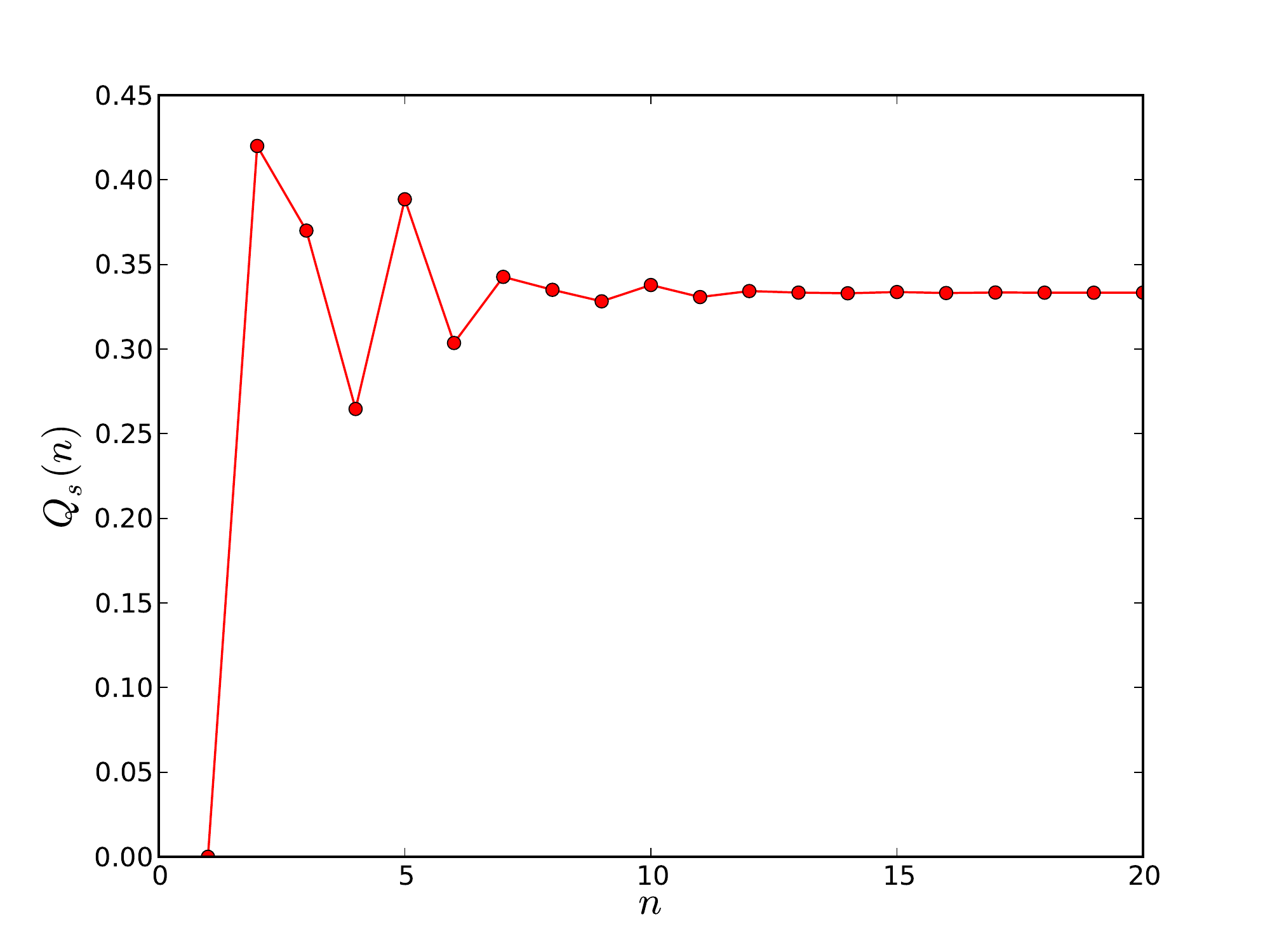}
\end{center}
\caption{$\Qsn$ vs. $n$ for $q=0.3$ the IID Process.}
\label{fig:BernoulliQs0p3q}
\end{figure}

Figures~\ref{fig:BernoulliQs0p1q} and~\ref{fig:BernoulliQs0p3q} show $\Qsn$
versus $n$ for the IID Process with $q=0.1$ and $q=0.3$, respectively, as
computed from Eq.~\eqref{eq:generalIIDQs}. In each case the CFs decay to an
asymptotic value of $1/3$, although this decay is faster for $q=0.3$. This is
not surprising, as one interpretation for the IID Process with $q=0.1$ is that
of a 3C$^+$ crystal interspersed with 10\% random deformation faults.

\subsection{Random Growth and Deformation Faults in Layered 3C and 2H CPSs: The RGDF Process}
\label{GrowthDeformationStacking}

\begin{figure}
\begin{center}
\includegraphics[width=0.50\textwidth]{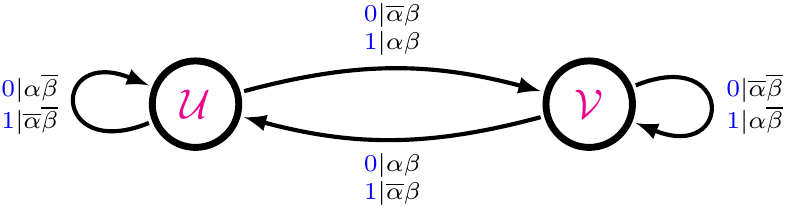}
\end{center}
\caption{RGDF Process, first proposed by Estevez~\etal~\cite{Este08a} and
  adapted here from Panel (c) of their Fig. (2). There is a slight change in
  notation. We relabeled the states given as `f' and `b' by
  Estevez~\etal~\cite{Este08a} as `$\mathcal{U}$' and `$\mathcal{V}$' and,
  instead of drawing an arc for each of the possible eight transitions, we took
  advantage of the multiple transitions between the same states and labeled
  each arc with two transitions. There is, of course, no change in meaning;
  this instead provides for slightly tidier illustration. Additionally, we
  correct a typographical error in Estevez~\etal~\cite{Este08a} when we
  relabel the transition ${\rm{b}} \xrightarrow{0|\overline{\alpha}\beta}
  {\rm{b}}$ with $\mathcal{V} \xrightarrow{0|\overline{\alpha}\overline{\beta}}
  \mathcal{V}$.
  }
\label{fig:EstevezHagg}
\end{figure}

Estevez \etal~\cite{Este08a} recently showed that simultaneous random
growth and deformation SFs in 2H and 3C CPSs can be modeled for all values of
the fault parameters by a simple HMM, and this is shown in
Fig.~\ref{fig:EstevezHagg}.  We refer to this process as the \emph{Random
Growth and Deformation Faults} (RGDF) Process.\footnote{ Estevez
\etal~\cite{Este08a} give a thorough and detailed discussion of the RGDF
process, and readers interested in a comprehensive motivation and derivation of
the RGDF process are urged to consult that reference.}  As has become
convention~\cite{Warr69a,Este08a}, $\alpha$ refers to deformation faulting and
$\beta$ refers to growth faults.

The HMM describing the RGDF Process is unlike any of the others considered here
in that on emission of a symbol from a state, the successor state is {\emph
{not}} uniquely specified. For example, $\mathcal{U} \xrightarrow{0}
\mathcal{U}$ and $\mathcal{U} \xrightarrow{0} \mathcal{V}$; \ie, being in state
$\mathcal{U}$ and emitting a 0 does not uniquely determine the next state. Such
a representations were previously called {\emph
{nondeterministic}}~\cite{Hopc79a}, but to avoid a conflict in terminology we
prefer the term {\emph {nonunifilar}}~\cite{Ephr02a,Elli09a}. Since \eMs\ are
unifilar~\cite{Crut89a,Shal01b}, the HMM representing the RGDF model is not an
\eM. Nonetheless, the techniques we have developed are applicable: CFs do not
require unifilar HMMs for their calculations, as do other properties such as
the entropy density.

Inspecting Fig.~\ref{fig:EstevezHagg}, the RGDF \MachineHagg's TMs are seen to
be (Eqs. (1) and (2) of Estevez \etal~\cite{Este08a}):
\begin{align*}
  \begin{array}{r@{\mskip\thickmuskip}l}
     \HaggTzero & = 
         \begin{bmatrix}
         \alpha \overline{\beta} & \overline{\alpha} \beta \\
         \alpha \beta & \overline{\alpha} \overline{\beta}
         \end{bmatrix} 
  \end{array} 
{\rm {and}}
  \begin{array}{r@{\mskip\thickmuskip}l}
      \HaggTone & = 
          \begin{bmatrix}
          \overline{\alpha} \overline{\beta} & \alpha \beta \\
          \overline{\alpha} \beta & \alpha \overline{\beta}
    \end{bmatrix} , 
  \end{array}
\end{align*}
where $\alpha \in [0, 1]$ and $\overline{\alpha} \equiv 1 - \alpha$, such that
$\alpha + \overline{\alpha} = 1$, and $\beta \in [0, 1]$ and $\overline{\beta}
\equiv 1 - \beta$, such that $\beta + \overline{\beta} = 1$.  There are eight
SSCs and, if at least one of them has $W^{\rm {SSC}} \neq 0$, the \Hagg-machine
is mixing. The self-state transitions each generate a nonvanishing $W^{\rm
{SSC}}$, so for the \MachineHagg\ to be nonmixing, these transitions must be
absent. Indeed, there are only two SSCs that have vanishing winding numbers,
and these are $[\mathcal{U}_{(0)}\mathcal{V}_{(1)}]$ and
$[\mathcal{U}_{(1)}\mathcal{V}_{(0)}]$.  These, and only these, SSCs can exist
if $\overline{\beta} = 0$ and $\alpha \in \{0,1\}$. Thus, the \MachineHagg\ is
nonmixing only for the parameter settings $\beta = 1$ and $\alpha \in \{0,1\}$,
which corresponds to the 2H crystal structure.

From the \Hagg-machine, we obtain the corresponding TMs of the $ABC$-machine for 
$\alpha, \beta \in (0, 1)$ by the rote expansion method (\S \ref{AlgebraicExpansion}): 
\begin{align}
\abcTA & = 
    \begin{bmatrix} \nonumber
       0 & 0 & 0 & 0 & 0 & 0 \\
       \alpha \overline{\beta} & 0 & 0 & \overline{\alpha} \beta & 0 & 0 \\
       \overline{\alpha} \overline{\beta} & 0 & 0 & \alpha \beta & 0 & 0 \\
       0 & 0 & 0 & 0 & 0 & 0 \\
       \alpha \beta & 0 & 0 & \overline{\alpha} \overline{\beta} & 0 & 0 \\
       \overline{\alpha} \beta & 0 & 0 & \alpha \overline{\beta} & 0 & 0 \\
    \end{bmatrix} , \\ \vspace{.05in}
\abcTB & = 
    \begin{bmatrix} \nonumber
        0 & \overline{\alpha} \overline{\beta} & 0 & 0 & \alpha \beta & 0 \\
        0 & 0 & 0 & 0 & 0 & 0 \\
        0 & \alpha \overline{\beta} & 0 & 0 & \overline{\alpha} \beta & 0 \\
        0 & \overline{\alpha} \beta & 0 & 0 & \alpha \overline{\beta} & 0 \\
        0 & 0 & 0 & 0 & 0 & 0 \\
        0 & \alpha \beta & 0 & 0 & \overline{\alpha} \overline{\beta} & 0 \\
    \end{bmatrix} ,
\intertext{and}
\abcTC & = 
    \begin{bmatrix} \nonumber
        0 & 0 & \alpha \overline{\beta} & 0 & 0 & \overline{\alpha} \beta \\
        0 & 0 & \overline{\alpha} \overline{\beta} & 0 & 0 & \alpha \beta \\
        0 & 0 & 0 & 0 & 0 & 0 \\
        0 & 0 & \alpha \beta & 0 & 0 & \overline{\alpha} \overline{\beta} \\
        0 & 0 & \overline{\alpha} \beta & 0 & 0 & \alpha \overline{\beta} \\
        0 & 0 & 0 & 0 & 0 & 0 \\
    \end{bmatrix} , 
\end{align}
and the orientation-agnostic state-to-state TM:
\begin{align}
\abcT & = \abcTA + \abcTB + \abcTC .   \nonumber
\end{align}
Explicitly, we have: 
\begin{align}
\abcT & = 
    \begin{bmatrix}   \nonumber
        0 & \overline{\alpha} \overline{\beta} & \alpha \overline{\beta} & 0 & \alpha \beta & \overline{\alpha} \beta \\
        \alpha \overline{\beta} & 0 & \overline{\alpha} \overline{\beta} & \overline{\alpha} \beta & 0 & \alpha \beta \\
        \overline{\alpha} \overline{\beta} & \alpha \overline{\beta} & 0 & \alpha \beta & \overline{\alpha} \beta & 0 \\
        0 & \overline{\alpha} \beta & \alpha \beta & 0 & \alpha \overline{\beta} & \overline{\alpha} \overline{\beta} \\
        \alpha \beta & 0 & \overline{\alpha} \beta & \overline{\alpha} \overline{\beta} & 0 & \alpha \overline{\beta} \\
        \overline{\alpha} \beta & \alpha \beta & 0 & \alpha \overline{\beta} & \overline{\alpha} \overline{\beta} & 0 \\
    \end{bmatrix} .
\end{align}

$\abcT$'s eigenvalues satisfy det$(\abcT - \lambda \IdentMat) = 0$. Here, 
with $a \equiv \alpha \beta$, $b \equiv \alpha \overline{\beta}$, $c \equiv \overline{\alpha} \beta$, and $d \equiv \overline{\alpha} \overline{\beta}$, 
we have:
\begin{align*}
\text{det}&(\abcT - \lambda \IdentMat) \\
& = 
%\bigl[ \lambda^2 - 2 \lambda (b+d) + (b+d)^2 - (a+c)^2  \bigr] \\
\bigl[ \bigl( \lambda -  (b+d) \bigr)^2 - (a+c)^2  \bigr] \\
& \quad \times   \bigl[ \lambda^2 + \lambda (b+d) + ac - bd - a^2 - c^2 + b^2 + d^2  \bigr]^2 \\
& = 0 , 
\end{align*}
from which we obtain the eigenvalues:
$\lambda = b+d \pm (a+c)$
and 
$\lambda = \OneHalf(b+d) \pm \OneHalf\left[ 4(a+c)^2 - 3(b+d)^2 + 12(bd - ac) \right]^{\OneHalf} $. 
To get back to $\alpha$s and $\beta$s, we note that 
$a+c = \beta$, $b+d = \overline{\beta}$, 
$ac = \beta^2 \alpha \overline{\alpha}$, and $bd = \overline{\beta}^2 \alpha \overline{\alpha}$. 
It also follows that $b+d+a+c = 1$, $b+d - (a+c) = \overline{\beta} - \beta = 1 - 2\beta$, and $bd-ac = \alpha \overline{\alpha}(\overline{\beta}^2 - \beta^2) = \alpha \overline{\alpha}(1 - 2\beta) = \alpha \overline{\alpha}(\overline{\beta} - \beta) $. 
Hence, after simplification, the set of $\abcT$'s eigenvalues can be written as:
\begin{align}
\Lambda_{\abcT} 
%\label{eq:EstevezEigsAsAlphasBetasWBars}
%&= \left\{ 1, \, \overline{\beta} - \beta, \, \tfrac{1}{2} \overline{\beta} \pm \tfrac{1}{2} \left[ 4\beta^2 - 3\overline{\beta}^2 + 12 \alpha \overline{\alpha}(\overline{\beta} - \beta) \right]^{\frac{1}{2} }  \right\} \\
\label{eq:EstevezEigsWSigma}
& = \left\{ 1, \, 1 - 2\beta, \, -\tfrac{1}{2} (1 - \beta) \pm \tfrac{1}{2} \sqrt{\sigma}  \right\} , 
\end{align}
with 
\begin{align}
\label{eq:ExpansionOfSigma}
\sigma &\equiv
4\beta^2 - 3\overline{\beta}^2 + 12 \alpha \overline{\alpha}(\overline{\beta} - \beta) \\
& = -3 + 12\alpha + 6\beta - 12\alpha^2 + \beta^2 - 24\alpha \beta + 24\alpha^2 \beta .
\end{align}

Except for measure-zero submanifolds along which the eigenvalues become extra
degenerate, throughout the parameter range the eigenvalues' algebraic
multiplicities are: $a_1 = 1$, $a_{1-2\beta} = 1$, $a_{- \tfrac{1}{2} (1 -
\beta + \sqrt{\sigma})} = 2$, and $a_{- \tfrac{1}{2} (1 - \beta -
\sqrt{\sigma})} = 2$.
Moreover, the \emph{index} of all eigenvalues is 1 except along $\sigma = 0$.

Immediately from the eigenvalues and their corresponding indices, we know all
possible characteristic modes of CF decay. All that remains is to find the
contributing amplitude of each characteristic mode. For comparison, note that
our $\sigma$ turns out to be equivalent to the all-important $-s^2$ term
defined in Eq.~(28) of Estevez \etal~\cite{Este08a}. 

Eqs.~\eqref{eq:EstevezEigsWSigma} and \eqref{eq:ExpansionOfSigma} reveal an
obvious symmetry between $\alpha$ and $\overline{\alpha}$ that is \emph{not}
present between $\beta$ and $\overline{\beta}$. In particular, $\abcT$'s
eigenvalues are invariant under exchange of $\alpha$ and
$\overline{\alpha}$---the CFs will decay in the same manner for $\alpha$-values
symmetric about $1/2$. There is no such symmetry between $\beta$ and
$\overline{\beta}$. Parameter space organization is seen nicely in Panel (c) of
Fig.~6 from Estevez \etal~\cite{Este08a}. Importantly, in that figure
$\sigma = 0$ should be seen as the critical line organizing a phase transition in
parameter space. Here, we will show that the $\sigma = 0$ line actually
corresponds to nondiagonalizability of the TM and, thus, to 
the qualitatively different polynomial
behavior in the decay of the CFs predicted by our Eq.~\eqref{eq:GenCFdecomp}.
%and typical of critical phenomena at the boundaries of phase transitions.

Note that since $\abcT$ is doubly-stochastic (\ie, all rows sum to one
\emph{and} all columns sum to one), the all-ones vector is not only the right
eigenvector associated with the eigenvalue of unity, but also the left
eigenvector associated with unity.  Moreover, since the stationary distribution
$\bra{\Dist}$ is the left eigenvector associated with unity (recall that $\bra
\Dist \abcT = \bra \Dist$), the stationary distribution is the uniform
distribution: $\bra \Dist = \tfrac{1}{6} \begin{bmatrix} 1&1&1&1&1&1
\end{bmatrix}$, \ie, $\bra \Dist = \tfrac{1}{6}
\DistOneBackward$, for $\alpha, \beta \in (0, 1)$.  Hence, throughout this
range, the projection operator associated with unity is $\abcT_1 = \tfrac{1}{6}
\DistOne \DistOneBackward$.

It is interesting to note that the eigenvalue of $1-2\beta$ is associated with
the decay of out-of-equilibrium probability density between the \Hagg\ states
of $\mathcal{U}$ and $\mathcal{V}$---or at least between the $ABC$-state
clusters into which each of the \Hagg\ states have split.  Indeed, from the
\Hagg\ machine: $\Lambda_{\HaggT} = \{ 1, \, 1-2\beta \}$. So, questions about
the relative occupations of the \Hagg\ states themselves are questions invoking
the $1-2\beta$ projection operator. However, due to the antisymmetry of output
orientations emitted from each of these \Hagg\ states, the $1-2\beta$
eigenvalue will not make any direct contribution towards answering questions
about the process's output statistics. Specifically, $\Braket{\abcT_{1-2\beta}
^{\xi(\Alphabet)}} = 0$ for all $\xi \in \{\rm{c}, \rm{a}, \rm{s} \}$. 
Since $a_{1-2\beta} = 1$, the projection operator is simply the matrix
product of the right and left eigenvectors associated with $1-2\beta$. With
proper normalization, we have:  
\begin{align}
\abcT_{1-2\beta} & = 
\tfrac{1}{6} 
\ket{\boldsymbol{1-2\beta}} \bra{\boldsymbol{1-2\beta}}  \nonumber
\end{align}
with $\ket{\boldsymbol{1-2\beta}} = \begin{bmatrix} 1&1&1&-1&-1&-1 \end{bmatrix} ^\top$ 
and $\bra{\boldsymbol{1-2\beta}} =  \begin{bmatrix} 1&1&1&-1&-1&-1
\end{bmatrix}$ where $\top$ denotes matrix transposition. Then, one can easily
check via Eq.~\eqref{eq:DiagonalizableCorrelationAmplitudes} that indeed
$\Braket {\abcT_{1-2\beta}^{\xi(\Alphabet)}} = 0$ for all $\xi \in \{\rm{c}, \rm{a}, \rm{s}
\}$. 

%\alert{The following (and just above) regarding the diagonalizability for $\sigma \neq 0$ is unfounded.  It was based on an assumption drawn from the Estevez paper and is not guaranteed!
%Find the general formulae, and only use the diagonalizable results if the diagonalizability has been proven!}

To obtain an explicit expression for the CFs, we 
must obtain the remaining projection operators.  We can
always use Eq.~\eqref{eq:GeneralProjOpEqn}. However, to draw attention to
useful techniques, % applicable in special cases, 
we will break the remaining
analysis into two parts: one for $\sigma = 0$ and the other for $\sigma \neq
0$. In particular, for the case of $\sigma = 0$, we show that
nondiagonalizablity need not make the problem harder than the diagonalizable
case.

\subsubsection{$\sigma=0$:}

As mentioned earlier, the $\sigma = 0$ line is the critical line that organizes
a phase transition in the ML ordering. We also find that $\abcT$ is
nondiagonalizable only along the $\sigma = 0$ submanifold. For $\sigma = 0$,
the $\tfrac{1}{2} (1 - \beta) \pm \tfrac{1}{2} \sqrt{\sigma}$ eigenvalues of
Eq.~\eqref{eq:EstevezEigsWSigma} collapse to a single eigenvalue so that the
set of eigenvalues reduces to: $\left. \Lambda_{\abcT} \right|_{\sigma = 0} =
\left\{ 1, \, 1 - 2\beta, \, -\tfrac{1}{2} (1 - \beta) \right\} $ with
corresponding indices: $\nu_1 = 1$, $\nu_{1-2\beta} = 1$, and $\nu_{-
\overline{\beta}/2} = 2$.

In this case, the projection operators are simple to obtain. 
As in the general case, we have:
\begin{align}
\abcT_1 &= 
\tfrac{1}{6} \DistOne \DistOneBackward  \nonumber \\ 
& = 
\tfrac{1}{6} 
    \begin{bmatrix} \nonumber
    1 & 1 & 1 & 1 & 1 & 1 \\
    1 & 1 & 1 & 1 & 1 & 1 \\
    1 & 1 & 1 & 1 & 1 & 1 \\
    1 & 1 & 1 & 1 & 1 & 1 \\
    1 & 1 & 1 & 1 & 1 & 1 \\
    1 & 1 & 1 & 1 & 1 & 1 
    \end{bmatrix}
\end{align}
and 
\begin{align}
\abcT_{1-2\beta} & = 
\tfrac{1}{6} 
\ket{\boldsymbol{1-2\beta}} \bra{\boldsymbol{1-2\beta}}   \nonumber \\
& = 
\tfrac{1}{6} 
    \begin{bmatrix} \nonumber 
    1 & 1 & 1 & -1 & -1 & -1 \\
    1 & 1 & 1 & -1 & -1 & -1 \\
    1 & 1 & 1 & -1 & -1 & -1 \\
    -1 & -1 & -1 & 1 & 1 & 1 \\
    -1 & -1 & -1 & 1 & 1 & 1 \\
    -1 & -1 & -1 & 1 & 1 & 1 
    \end{bmatrix} .
\end{align}
Recall that the projection operators sum to the identity: $\IdentMat =
\sum_{\lambda \in \Lambda_{\abcT}} \abcT_\lambda = \abcT_1 + \abcT_{1-2\beta} +
\abcT_{-\overline{\beta}/2} $. And so, it is easy to obtain the
remaining projection operator:
\begin{align}
\abcT_{-\overline{\beta}/2} & = 
\IdentMat - \abcT_1 - \abcT_{1-2\beta} \nonumber \\ 
& = 
\tfrac{1}{3} 
    \begin{bmatrix} \nonumber 
    2 & -1 & -1 & 0 & 0 & 0 \\
    -1 & 2 & -1 & 0 & 0 & 0 \\
    -1 & -1 & 2 & 0 & 0 & 0 \\
    0 & 0 & 0 & 2 & -1 & -1 \\
    0 & 0 & 0 & -1 & 2 & -1 \\
    0 & 0 & 0 & -1 & -1 & 2 
    \end{bmatrix} .
\end{align} 

Note that $3 \bra \Dist \abcTA = \tfrac{1}{2} \DistOneBackward \abcTA = 
\tfrac{1}{2} \begin{bmatrix} 1 & 0 & 0 & 1 & 0 & 0 \end{bmatrix}$ and 
that:
\begin{align}
\abcTA \DistOne = &
\begin{bmatrix}
       0 \\
       \alpha \overline{\beta} + \overline{\alpha} \beta  \\
       \alpha \beta + \overline{\alpha} \overline{\beta}  \\
       0 \\
       \alpha \beta + \overline{\alpha} \overline{\beta}  \\
       \alpha \overline{\beta} + \overline{\alpha} \beta  \\
    \end{bmatrix} , 
\abcTB \DistOne  = 
    \begin{bmatrix}
         \alpha \beta + \overline{\alpha} \overline{\beta}  \\
        0  \\
        \alpha \overline{\beta} + \overline{\alpha} \beta  \\
        \alpha \overline{\beta} + \overline{\alpha} \beta  \\
        0  \\
        \alpha \beta + \overline{\alpha} \overline{\beta}  \\
    \end{bmatrix} , \nonumber \\
\text{ and } & \; 
\abcTC \DistOne = 
    \begin{bmatrix}
        \alpha \overline{\beta} + \overline{\alpha} \beta \\
        \alpha \beta + \overline{\alpha} \overline{\beta}  \\
        0  \\
        \alpha \beta + \overline{\alpha} \overline{\beta} \\
        \alpha \overline{\beta} + \overline{\alpha} \beta  \\
        0 \\
    \end{bmatrix} ~. \nonumber
\end{align}
Then, 
according to Eq.\ \eqref{eq:GenCFdecomp}, with 
$\Braket{\abcT_1 ^{\xi(\Alphabet)}}  = \tfrac{1}{3}$, 
$\Braket{\abcT_{1-2\beta} ^{\xi(\Alphabet)}} = 0$, 
$\Braket{\abcT_{-\overline{\beta}/2} ^{s(\Alphabet)}} = -\tfrac{1}{3}$, 
$\Braket{\abcT_{-\overline{\beta}/2} ^{c(\Alphabet)}} = 
\Braket{\abcT_{-\overline{\beta}/2} ^{a(\Alphabet)}} = \tfrac{1}{6}$,  
$\Braket{\abcT_{-\overline{\beta}/2, 1} ^{s(\Alphabet)}} = 
\tfrac{1}{6} (\sigma + \beta - \beta^2) = \tfrac{1}{6} \beta \overline{\beta}$, and 
$\Braket{\abcT_{-\overline{\beta}/2, 1} ^{c(\Alphabet)}} = 
\Braket{\abcT_{-\overline{\beta}/2, 1} ^{a(\Alphabet)}} = 
-\tfrac{1}{12} (\sigma + \beta - \beta^2)  = -\tfrac{1}{12} \beta \overline{\beta}$, the CFs are: 
\begin{align}
\Qxin & = 
  \sum_{\lambda \in \Lambda_{\abcT}} 
  \sum_{m=0}^{\nu_\lambda - 1}  
  \CFBraKetm \binom{n-1}{m}  \lambda^{n-m-1} \nonumber \\ 
& = \Braket{\abcT_1 ^{\xi(\Alphabet)}} + 
  \sum_{m=0}^{1}  
  \Braket{\abcT_{-\overline{\beta}/2, m} ^{\xi(\Alphabet)}}  \binom{n-1}{m}  {\left(-\overline{\beta}/2 \right)}^{n-m-1} \nonumber \\ 
%& = \tfrac{1}{3} + \left[ \Braket{\abcT_{-\overline{\beta}/2} ^{\xi(\Alphabet)}} - \frac{2 (n-1)}{\overline{\beta}} \Braket{\abcT_{-\overline{\beta}/2, 1} ^{\xi(\Alphabet)}} \right]  {\left(-\overline{\beta}/2 \right)}^{n-1} \nonumber 
& = \tfrac{1}{3} + \left[ \Braket{\abcT_{-\overline{\beta}/2} ^{\xi(\Alphabet)}} - \frac{2}{\overline{\beta}} \Braket{\abcT_{-\overline{\beta}/2, 1} ^{\xi(\Alphabet)}} (n-1) \right]  {\left(-\overline{\beta}/2 \right)}^{n-1} \nonumber 
  ~.
\end{align}
Specifically: 
\begin{align}
\Qsn & =  \tfrac{1}{3} \left[ 1 + 2 \left( 1 + \frac{\beta}{\overline{\beta}} \, n \right) {\left(-\overline{\beta}/2 \right)}^{n}  \right]~, 
\end{align}
and
\begin{align}
\Qcn = 
\Qan =  \tfrac{1}{3} \left[ 1 - \left( 1 + \frac{\beta}{\overline{\beta}} \, n \right) {\left(-\overline{\beta}/2 \right)}^{n} \right]~.
\end{align}

\subsubsection{$\sigma \neq 0$:}

For any value of $\sigma$, excluding of course $\sigma = 0$, we can obtain
the projection operators via Eq.~\eqref{eq:GeneralProjOpEqn}. 
In addition to those quoted above and, in terms of the former $\abcT_{-\overline{\beta}/2}$, the remaining projection operators turn out to be:
\begin{align*}
\abcT_{\frac{-\overline{\beta} \pm \sqrt{\sigma}}{2}} 
= \pm \frac{1}{\sqrt{\sigma}} \abcT_{-\overline{\beta}/2} 
\left[ \abcT + \left( \frac{\overline{\beta} \pm \sqrt{\sigma} }{2} \right) \IdentMat \right]~. 
\end{align*}

Since the $1-2\beta$ eigen-contribution is null and since:
\begin{align*}
\Braket{T_{1}^{\xi(\Alphabet)}} & = 1/3 ~,  \\
\Braket{T_{\frac{-\overline{\beta} \pm \sqrt{\sigma}}{2}}^{s(\Alphabet)}}
  & = \frac{1}{6} \left[ -1 \pm \left( \sqrt{\sigma}
  + \frac{\beta \overline{\beta}}{\sqrt{\sigma}} \right) \right] \\
  & = \pm \frac{1}{6} \left( 1 \mp \frac{\beta}{\sqrt{\sigma}} \right)
  \left( \sqrt{\sigma} \mp \overline{\beta} \right) ~, \text{~and} \\
\Braket{T_{\frac{-\overline{\beta} \pm \sqrt{\sigma}}{2}}^{c(\Alphabet)}}
  & = \Braket{T_{\frac{-\overline{\beta} \pm \sqrt{\sigma}}{2}}^{a(\Alphabet)}}
  \\
  & = \frac{1}{12} \left[ 1 \mp \left( \sqrt{\sigma} + \frac{\beta
  \overline{\beta}}{\sqrt{\sigma}} \right) \right] \\
  & = \mp \frac{1}{12} \left( 1 \mp \frac{\beta}{\sqrt{\sigma}} \right)
    \left( \sqrt{\sigma} \mp \overline{\beta} \right)
  ~,
\end{align*}
the CFs for $\sigma \neq 0$ are: 
\begin{align}
\Qxin &=  
\sum_{\lambda \in \Lambda_{\abcT}} 
\lambda^{n-1} 
\sum_{\symbolABC_{0} \in \AlphabetABC} \bra \Dist 
                 \TransMatABC^{[\symbolABC_{0}]} \abcT_\lambda
                \TransMatABC^{[\xoperator (\symbolABC_{0})]} \DistOne \nonumber \\
& = 
\tfrac{1}{3} + 
\sum_{\lambda \in \{ 
\frac{-\overline{\beta} \pm \sqrt{\sigma}}{2}
\}} 
 \CFBraKet  \lambda^{n-1} ~. 
\label{eq:RandomStackingQxinWith2ProjOps}
\end{align}
Specifically, for $\xi = s$:
\begin{align}
\Qsn = \tfrac{1}{3} 
&+ \tfrac{1}{6} \left( 1 - \tfrac{\beta}{\sqrt{\sigma}} \right) \left( \sqrt{\sigma} - \overline{\beta} \right)
\left( \tfrac{-\overline{\beta} + \sqrt{\sigma}}{2} \right)^{n-1} \nonumber \\
&- \tfrac{1}{6} \left( 1 + \tfrac{\beta}{\sqrt{\sigma}} \right) \left( \sqrt{\sigma} + \overline{\beta} \right)
\left( \tfrac{-\overline{\beta} - \sqrt{\sigma}}{2} \right)^{n-1} \\ 
\nonumber \\
= 
\tfrac{1}{3} \Bigl[ 1 + &
\left( 1 - \tfrac{\beta}{\sqrt{\sigma}} \right) \left( \tfrac{-\overline{\beta} + \sqrt{\sigma}}{2} \right)^{n} 
+ \left( 1 + \tfrac{\beta}{\sqrt{\sigma}} \right) \left( \tfrac{-\overline{\beta} - \sqrt{\sigma}}{2} \right)^{n}
\Bigr]~, \nonumber
\end{align}
and we recover Eq.~(29) of
Estevez~\etal~\cite{Este08a}.

\begin{figure}
\begin{center}
\includegraphics[width=0.5\textwidth]{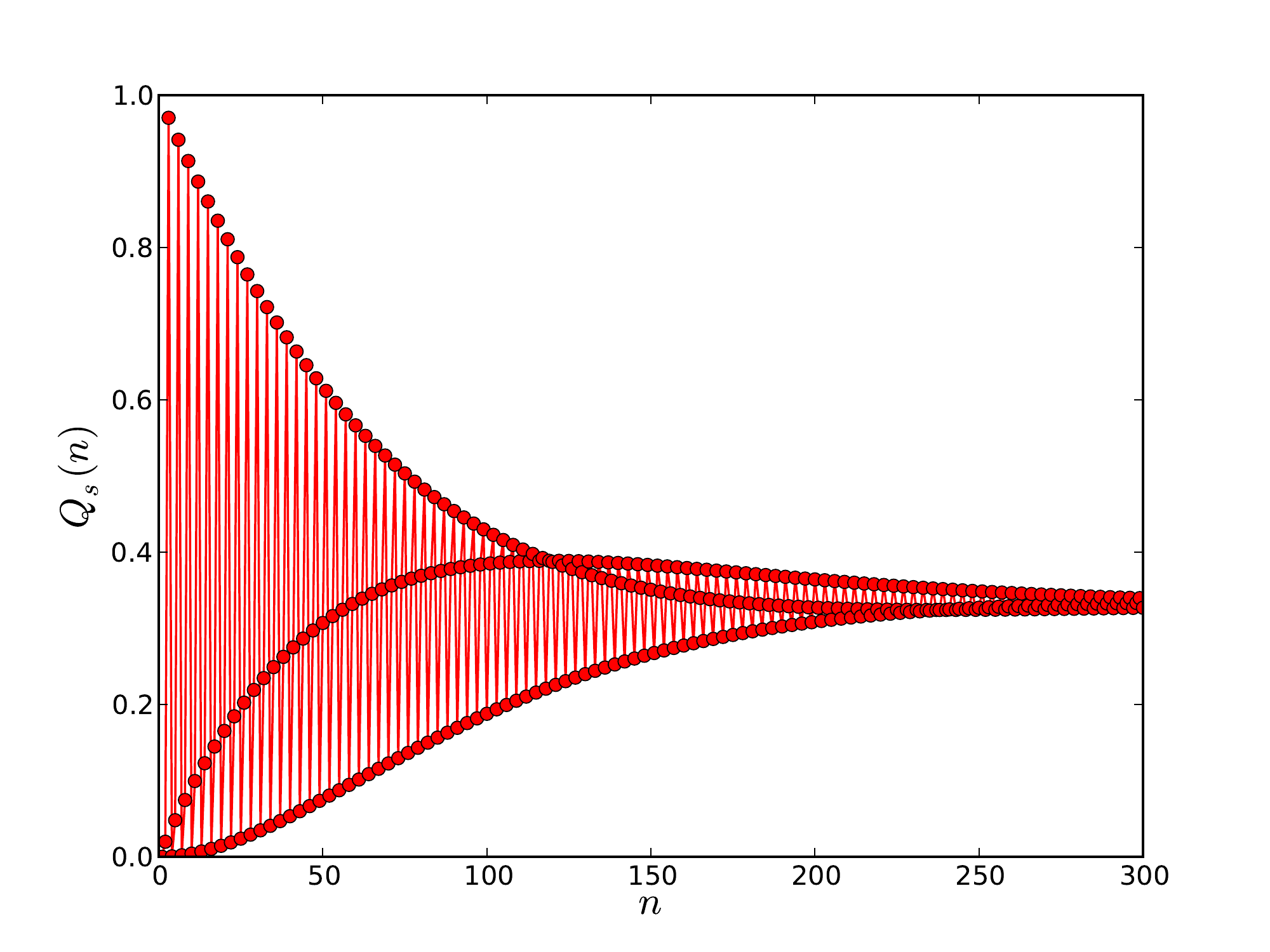}
\end{center}
\caption{$\Qsn$ vs. $n$ with $\alpha = 0.01$ and $\beta = 0$ for the RGDF
  Process. This should be compared to Panel (b) of Fig.~8 in
  Estevez~\etal~\cite{Este08a}. Although different means were used to make
  the calculations, they appear to be identical.
  }
\label{fig:Estevez_Qs_0p01a_0b}
\end{figure}

\begin{figure}
\begin{center}
\includegraphics[width=0.5\textwidth]{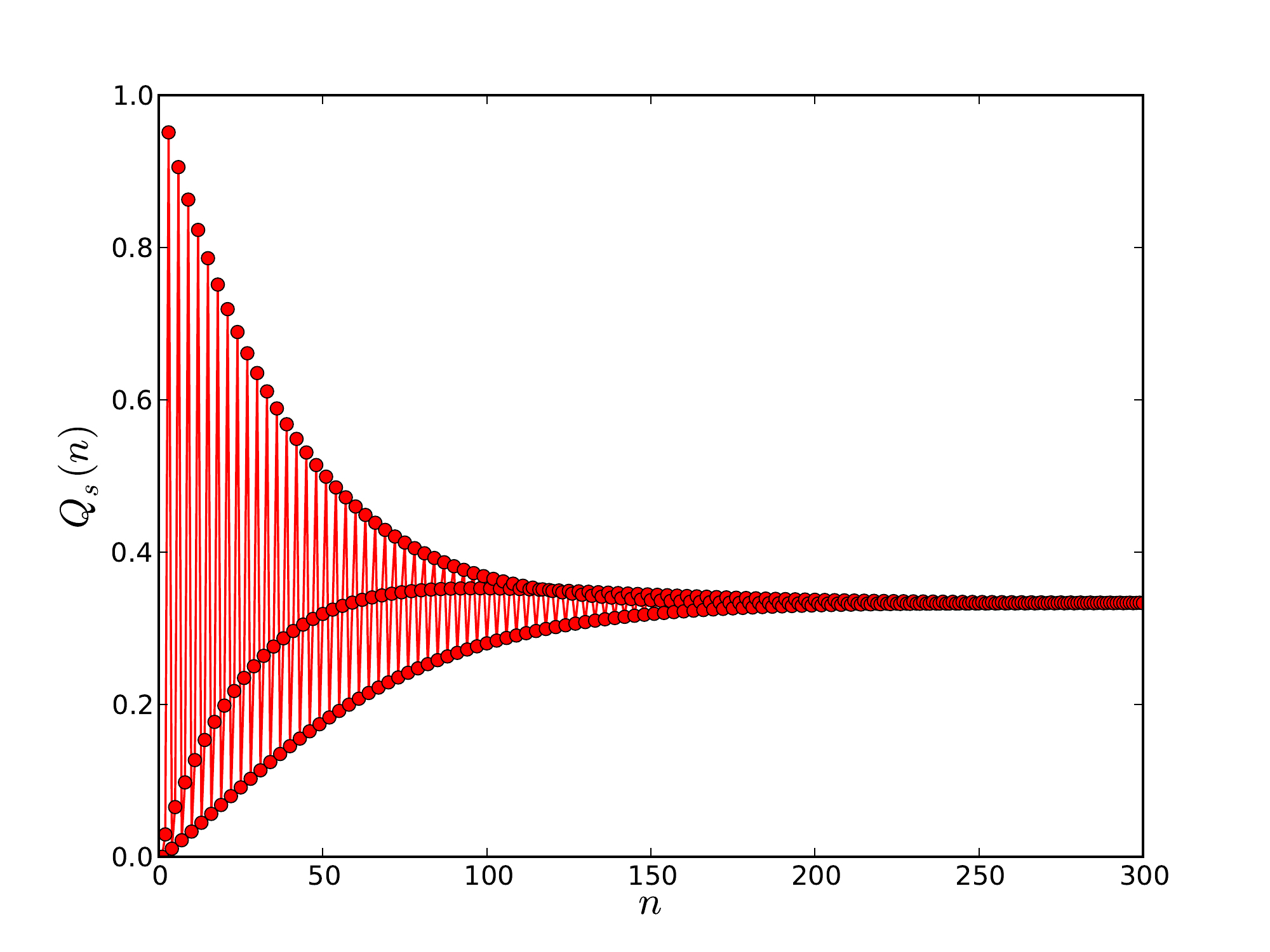}
\end{center}
\caption{$\Qsn$ vs. $n$ with $\alpha = 0.01$ and $\beta = 0.01$ for the RGDF
  Process. Comparison with Panel (d) of Fig.~8 in
  Estevez~\etal~\cite{Este08a} shows an identical result.
  }
\label{fig:Estevez_Qs_0p01a_0b01}
\end{figure}

\begin{figure}
\begin{center}
\includegraphics[width=0.5\textwidth]{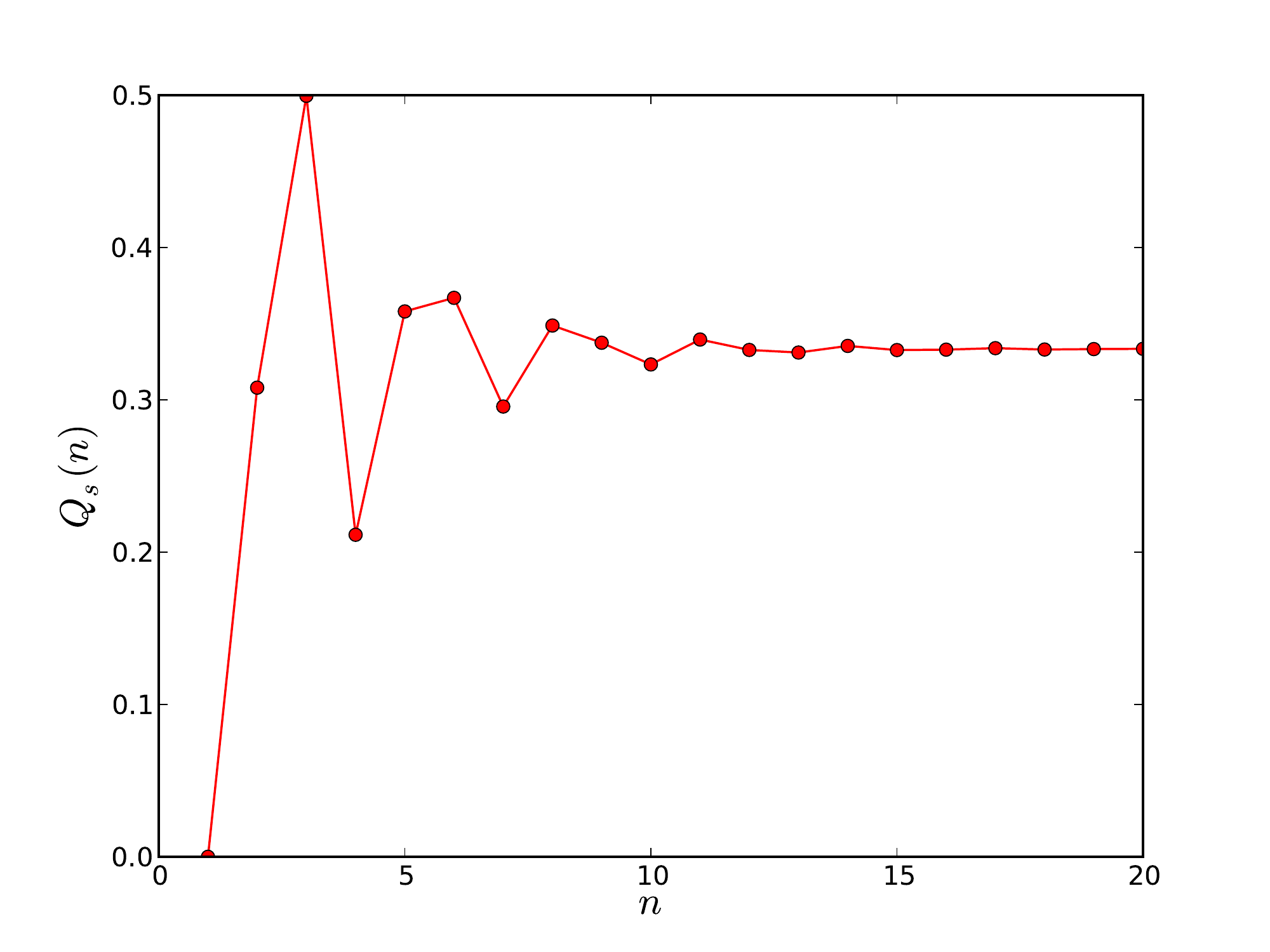}
\end{center}
\caption{$\Qsn$ vs. $n$ with $\alpha = 0.1$ and $\beta = 0.2$ for the RGDF
  Process.
  }
\label{fig:Estevez_Qs_0p1a_0b2}
\end{figure}

\begin{figure}
\begin{center}
\includegraphics[width=0.5\textwidth]{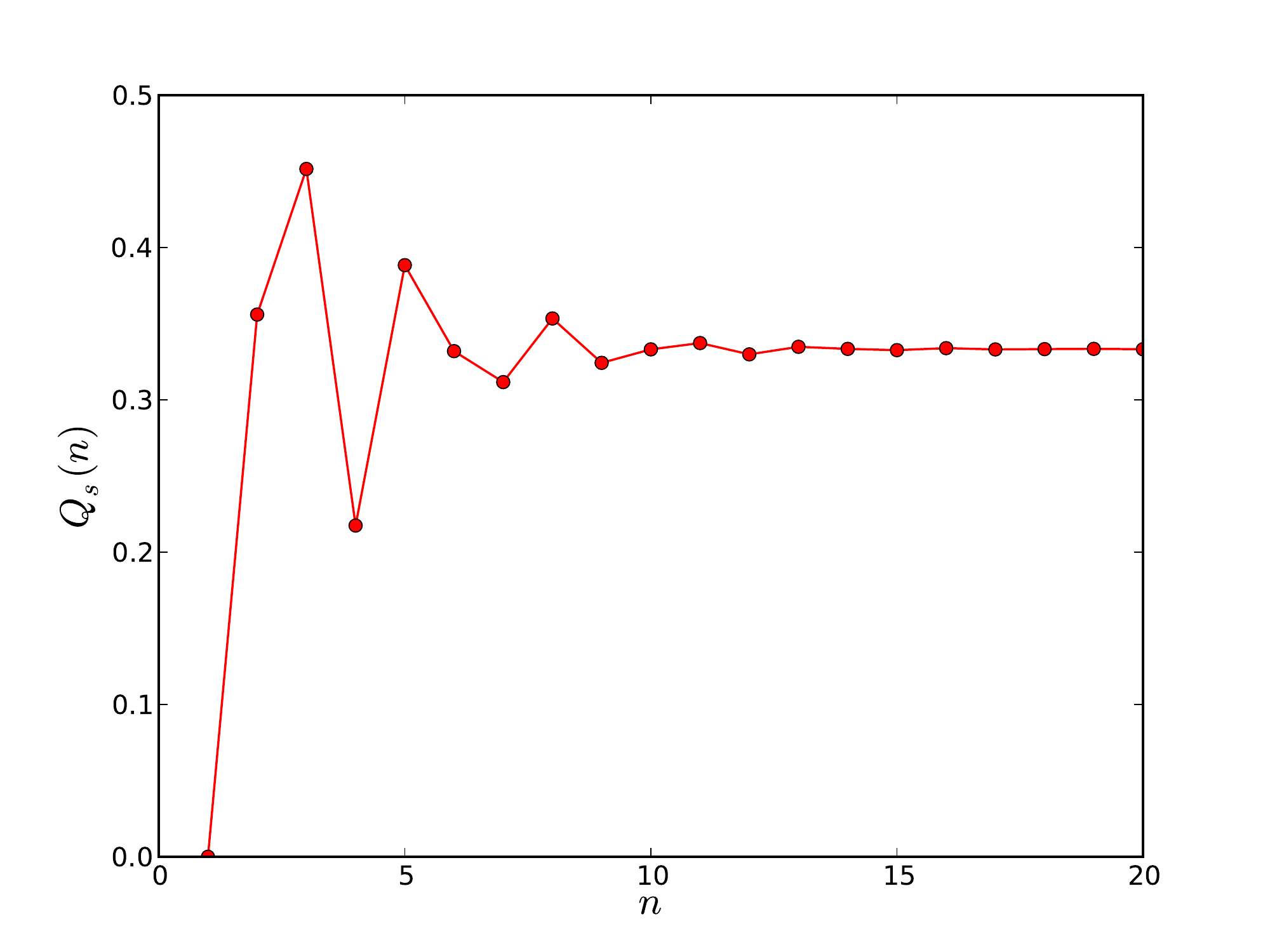}
\end{center}
\caption{$\Qsn$ vs. $n$ with $\alpha = 0.2$ and $\beta = 0.1$ for the RGDF
  Process.
  }
\label{fig:Estevez_Qs_0p2a_0b1}
\end{figure}

Estevez \etal~\cite{Este08a} recount the embarrassingly long list of recent
failures of previous attempts to analyze organization in RGDF-like processes.
These failures resulted from not obtaining all of the terms in the CFs, which in turn
stem primarily from not using a sufficiently clever ansatz in their
methods, together with not knowing how many terms there should be. In contrast,
even when casually observing the number of HMM states, our method gives
immediate knowledge of the number of terms. Our method is generally applicable
with straightforward steps to actually calculate all the terms once and for
all.

Figures~\ref{fig:Estevez_Qs_0p01a_0b},~\ref{fig:Estevez_Qs_0p01a_0b01},~\ref{fig:Estevez_Qs_0p1a_0b2}
and~\ref{fig:Estevez_Qs_0p2a_0b1} show plots of $\Qsn$ versus $n$ for the
RGDF Process at different values of $\alpha$ and $\beta$. The first two graphs,
Figs.~\ref{fig:Estevez_Qs_0p01a_0b} and~\ref{fig:Estevez_Qs_0p01a_0b01}, were
previously produced by Estevez~\etal~\cite{Este08a} and appear to be
identical to our results. The second pair of graphs for the IID Process,
Figs.~\ref{fig:Estevez_Qs_0p1a_0b2} and~\ref{fig:Estevez_Qs_0p2a_0b1} show the
behavior of the CFs for larger values of $\alpha$ and $\beta$, but with the
numerical values of each exchanged ($0.1 \Leftrightarrow 0.2$).  The CFs are
clearly sensitive to the kind of faulting present, as one would expect.
However, each does decay to $1/3$, as they must.

\subsection{Shockley--Frank Stacking Faults in 6H-SiC: The SFSF Process}
\label{ShockleyFrankStacking}

While promising as a material for next generation electronic components,
fabricating SiC crystals of a specified polytype remains challenging. Recently
Sun~\etal~\cite{Sun12a} reported experiments on 6H-SiC epilayers ($\sim 200
\, \mu$m thick) produced by the fast sublimation growth process at 1775 $^{\circ}$C.
Using high resolution transmission electron microscopy (HRTEM), they were able
to survey the kind and amount of particular SFs present. In the \Hagg\ notation 6H-SiC is 
specified by 000111, and this is written in the Zhdanov notation as (3,3)~\cite{Orti13a}. 
Thus, unfaulted 6H-SiC can be thought of as alternating blocks of size-three domains.
Ab initio super-cell calculations by
Iwata~\etal~\cite{Iwat03a} predicted that the Shockley defects (4,2),
(5,1), (9,3), and (10,2) should be present, with the (4,2) defect having the
lowest energy and, thus, it presumably should be the most common. Of these, however,
Sun~\etal~\cite{Sun12a} observed only the (9,3) defect (given there
as (3,9)) and, at that, only once. Instead, the most commonly observed
defects were (3,4), (3,5), (3,6), and (3,7), appearing nine, two, two, and three
times respectively, with isolated instances of other SF sequences.  They
postulated that combined Shockley--Frank defects~\cite{Hirt68a} could produce
these results. The (3,4) stacking sequences could be explained as external
Frank SFs, and the other observed faults could result from further Shockley
defects merging with these (3,4) SFs. We call this process the
\emph{Shockley-Frank Stacking Fault} (SFSF) Process.

\begin{figure}
\begin{center}
\includegraphics[width=0.5\textwidth]{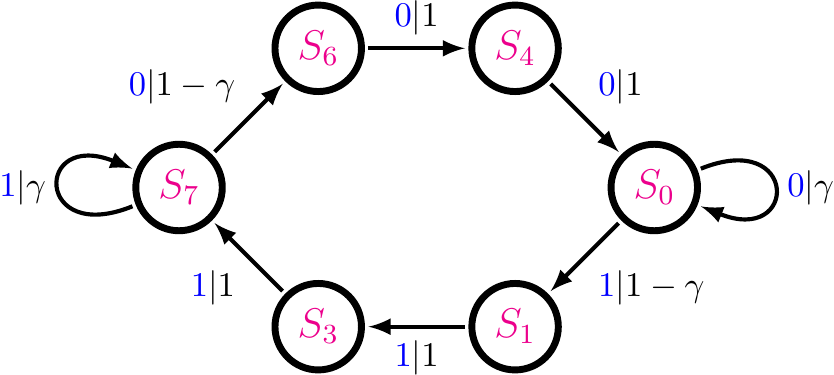}
\end{center}
\caption{\MachineHagg\ for the SFSF Process, inspired by the observations of
  Sun \etal~\cite{Sun12a}. We observe that there is one faulting parameter
  $\gamma \in [0,1]$ and three SSCs. Or, equivalently three CSCs, as this graph
  is also an \eM. The three SSCs are [$\sf{S}_7$], [$\sf{S}_0]$ and  [$\sf{S}_7
  \sf{S}_6 \sf{S}_4 \sf{S}_0 \sf{S}_1 \sf{S}_3$]. The latter we recognize as
  the 6H structure if $\gamma = 0$. For large values of $\gamma$, \ie, as
  $\gamma \to 1$, this process approaches a twinned 3C structure, although the
  faulting is {\emph {not}} random. The causal state architecture prevents the
  occurrence of domains of size-three or less.
  }
\label{fig:SunHaggMachine}
\end{figure}

Inspired by these observations, we ask what causal-state structure could
produce such stacking sequences.  We suggest that the \eM\ shown in
Fig.~\ref{fig:SunHaggMachine} is a potential candidate, with $\gamma \in [0,1]$
as the sole faulting parameter. (Here, we must insist that only a thorough
analysis, with significantly more HRTEM data or a DP, can properly reveal the
appropriate causal-state structure. The SFSF Process is given primarily to
illustrate our methods.) For weakly faulted crystals ($\gamma \approx 0$), as
seems to be the case here, there must be a CSC that gives the 6H structure, and we
see that the causal-state sequence [$\sf{S}_7 \sf{S}_6 \sf{S}_4 \sf{S}_0
\sf{S}_1 \sf{S}_3$] does that. Indeed, if the fault parameter $\gamma$ were
identically zero, then this \eM\ would give only the 6H structure.
Sun~\etal~\cite{Sun12a}'s observations suggest that deviations from 6H
structure occur (almost) always as {\emph {additions}} to the size-three 0 or 1
domains. The self-state transitions on $\sf{S}_7$ and $\sf{S}_0$ have just this
effect: After seeing three consecutive 1s (0s), with probability $\gamma$ the
current domain will increase in size to four.  And likewise, with probability
$\gamma$, size-four domains will increase to size-five domains. Thus, with
decreasing probability, the faults (3,4), (3,5) $\dots$ can be modeled by this
\eM. Notice that the causal state architecture prevents domains of any size
less that three, which is consistent with the bulk of the observations by
Sun~\etal~\cite{Sun12a}.\footnote{They did observe a single (3,2) sequence
(see their Table I), and the SFSF Process cannot reproduce that structure.
Additional causal states and/or transitions would be needed to accommodate this
additional stacking structure.  One obvious and simple modification that would
produce domains of size-two would be to allow the transitions $\sf{S}_3
\xrightarrow{0} \sf{S}_6$  and $\sf{S}_4 \xrightarrow{1} \sf{S}_1$ with some
small probability. However, in the interest maintaining a reasonably clear
example, we neglect this possibility.} Also, this \eM\ does predict (4,4)
sequences, which Sun~\etal~\cite{Sun12a} observed once. Thus,
qualitatively, and approximately quantitatively, the proposed \eM\ largely
reproduces the observations of Sun~\etal~\cite{Sun12a}.

We begin by identifying the SSCs on the HMM, the \eM\ shown in
Fig.~\ref{fig:SunHaggMachine}. We find that there are three, [$\sf{S}_7$],
[$\sf{S}_0]$ and  [$\sf{S}_7 \sf{S}_6 \sf{S}_4 \sf{S}_0 \sf{S}_1 \sf{S}_3$]. We
calculate the winding numbers to be $W^{[\sf{S}_7]} = 1$, $W^{[\sf{S}_0]} = 2$,
and $W^{[\sf{S}_7 \sf{S}_6 \sf{S}_4 \sf{S}_0 \sf{S}_1 \sf{S}_3]} = 0$. The
first two of these SSCs vanish if $\gamma = 0$, giving a nonmixing
\MachineHagg. Thus, for $\gamma \neq 0$ the \MachineHagg\ is mixing and we
proceed with the case of $\gamma \in (0,1]$.

By inspection we write down the two 6-by-6 TMs of the \MachineHagg\ as:
\begin{align}
\HaggTzero & = 
  \begin{bmatrix} \nonumber 
  \gamma & 0 & 0 & 0 & 0 & 0 \\
  0 & 0 & 0 & 0 & 0 & 0 \\
  0 & 0 & 0 & 0 & 0 & 0 \\ 
  0 & 0 & 0 & 0 & \overline{\gamma} & 0 \\ 
  0 & 0 & 0 & 0 & 0 & 1 \\ 
  1 & 0 & 0 & 0 & 0 & 0 \\ 
  \end{bmatrix}
\intertext{and:}
\HaggTone & = 
  \begin{bmatrix} \nonumber 
  0 & \overline{\gamma} & 0 & 0 & 0 & 0 \\
  0 & 0 & 1 & 0 & 0 & 0 \\
  0 & 0 & 0 & 1 & 0 & 0 \\ 
  0 & 0 & 0 & \gamma & 0 & 0 \\ 
  0 & 0 & 0 & 0 & 0 & 0 \\ 
  0 & 0 & 0 & 0 & 0 & 0 \\ 
  \end{bmatrix} ,
\end{align}
where the states are ordered $\sf{S}_0$, $\sf{S}_1$, $\sf{S}_3$, $\sf{S}_7$,
$\sf{S}_6$, and $\sf{S}_4$. The internal state TM is their sum:
\begin{align*}
\HaggT & = 
  \begin{bmatrix} \nonumber 
  \gamma & \overline{\gamma} & 0 & 0 & 0 & 0 \\
  0 & 0 & 1 & 0 & 0 & 0 \\
  0 & 0 & 0 & 1 & 0 & 0 \\ 
  0 & 0 & 0 & \gamma & \overline{\gamma} & 0 \\ 
  0 & 0 & 0 & 0 & 0 & 1 \\ 
  1 & 0 & 0 & 0 & 0 & 0 \\ 
  \end{bmatrix} .
\end{align*}
Since the six-state \MachineHagg\ generates an ($3 \times 6 = $) eighteen-state
\MachineABC, we do not explicitly write out the TMs of the \MachineABC.
Nevertheless, it is straightforward to expand the \MachineHagg\ to the
\MachineABC\ via the rote expansion method of \S\ref{AlgebraicExpansion}.  It
is also straightforward to apply Eq.~\eqref{eq:ThreeTimesBraKetSimplification}
to obtain the CFs as a function of the faulting parameter $\gamma$. To use
Eq.~\eqref{eq:ThreeTimesBraKetSimplification}, note that the stationary
distribution over the \MachineABC\ can be obtained via
Eq.~\eqref{eq:PiExpansion} with:
\begin{align*}
\bra{\Dist_{\textrm{H}}} &= 
  \tfrac{1}{6 - 4 \gamma} 
  \begin{bmatrix}  1 & \overline{\gamma} & \overline{\gamma} & 1 & \overline{\gamma} & \overline{\gamma}  \end{bmatrix} 
\end{align*}
as the stationary distribution over the \MachineHagg. 

\begin{figure}
\begin{center}
\includegraphics[width=0.5\textwidth]{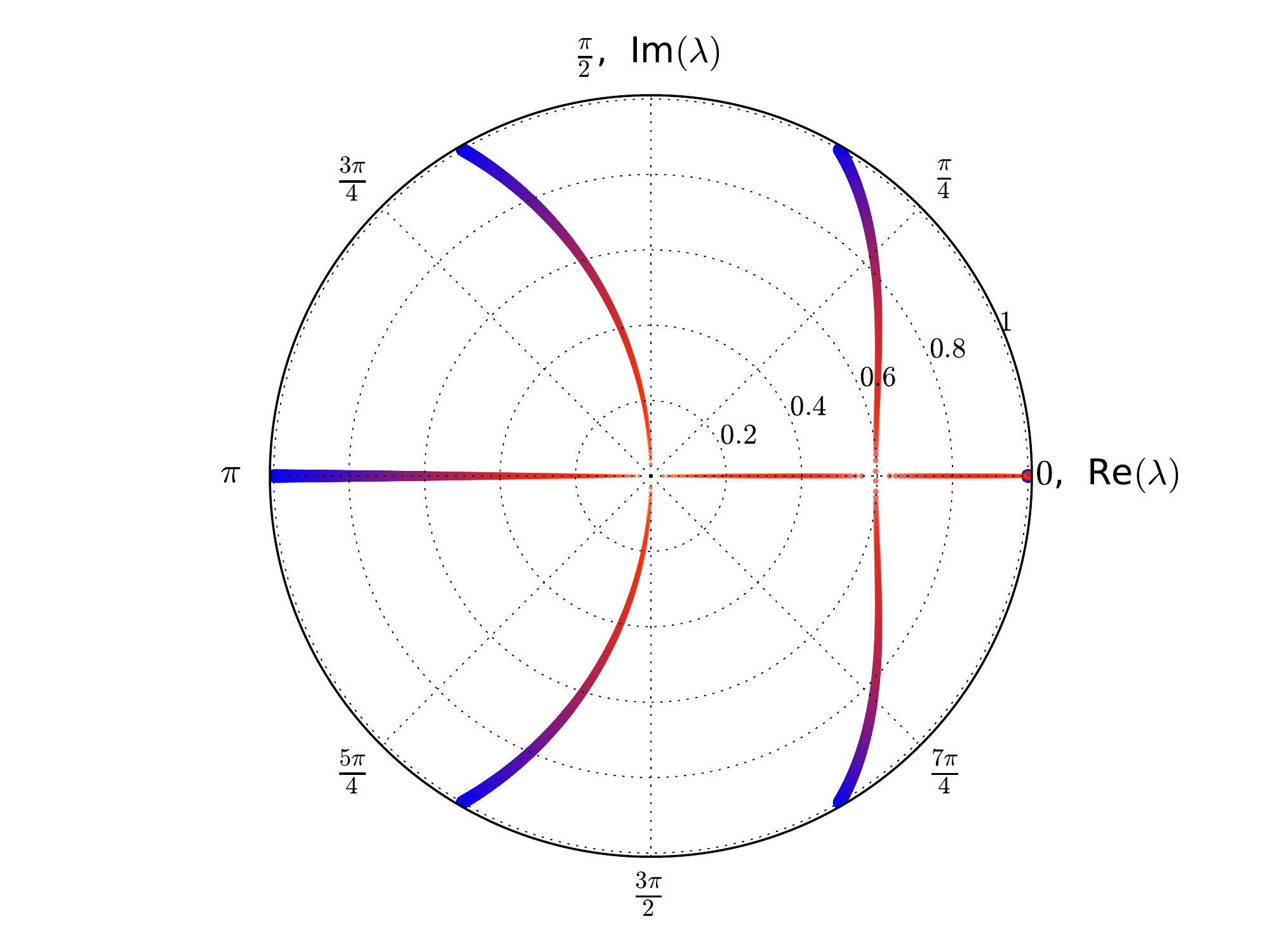}
\end{center}
\caption{The six eigenvalues of the \Hagg-machine as they evolve from
  $\gamma = 0$ (thickest blue markings) to $\gamma = 1$ (thinnest red
  markings). Note that the eigenvalues at $\gamma = 0$ are the six roots of
  unity. Unity is a persistent eigenvalue. Four of the eigenvalues approach 0
  as $\gamma \to 1$. Another of the eigenvalues approaches unity as $\gamma \to
  1$. The eigenvalues are nondegenerate throughout the parameter range except
  for the transformation event where the two eigenvalues on the right collide
  and scatter upon losing their imaginary parts.
  }
\label{fig:RealImagPartsHagg}
\end{figure}

The eigenvalues of the \Hagg-TM can be obtained as the solutions of
$\det(\HaggT - \lambda \IdentMat) = (\lambda - \gamma)^2 \lambda^4 -
\overline{\gamma}^2 = 0$.  These include $1$, $-\tfrac{1}{2} \overline{\gamma}
\pm \sqrt{\gamma^2 + 2\gamma - 3}$, and three other eigenvalues involving cube
roots.  Their values are plotted in the complex plane
Fig.~\ref{fig:RealImagPartsHagg} as we sweep through $\gamma$. 

\begin{figure}
\begin{center}
\includegraphics[width=0.5\textwidth]{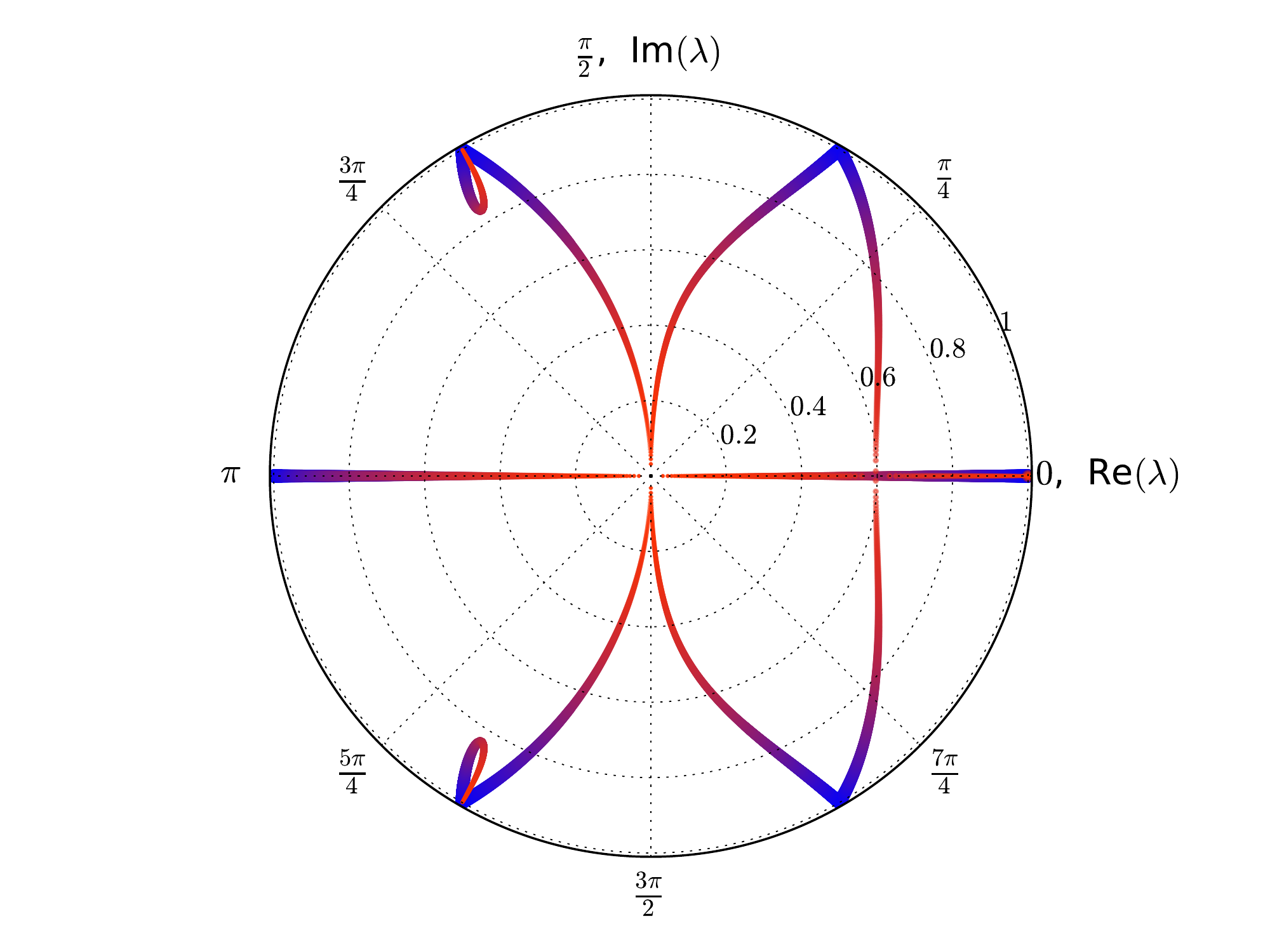}
\end{center}
\caption{The eighteen eigenvalues of the \MachineABC\ as they evolve from
  $\gamma =  0$ (thickest blue markings) to $\gamma = 1$ (thinnest red
  markings). Note that the eigenvalues at $\gamma = 0$ are still the six roots
  of unity. The new eigenvalues introduced via transformation to the
  \MachineABC\ all appear in degenerate (but diagonalizable) pairs. In terms of
  increasing $\gamma$, these include eigenvalues approaching zero from $\pm 1$,
  eigenvalues taking a left branch towards zero as they lose their imaginary
  parts, and eigenvalues looping away and back towards the nontrivial
  cube-roots of unity.
  }
\label{fig:RealImagPartsABC}
\end{figure}

The eigenvalues of the $ABC$-TM are similarly obtained as the solutions of
$\det(\abcT - \lambda \IdentMat) = 0$.  The real and imaginary parts of these
eigenvalues are plotted in Fig.\ \ref{fig:RealImagPartsABC}. Note that
$\Lambda_{\abcT}$ inherits $\Lambda_{\HaggT}$ as the backbone for its more
complex structure, just as $\Lambda_{\HaggT} \subseteq \Lambda_{\abcT}$ for all
of our previous examples. The eigenvalues in $\Lambda_{\abcT}$ are, of course,
those most directly responsible for the structure of the CFs.

\begin{figure}
\begin{center}
\includegraphics[width=0.5\textwidth]{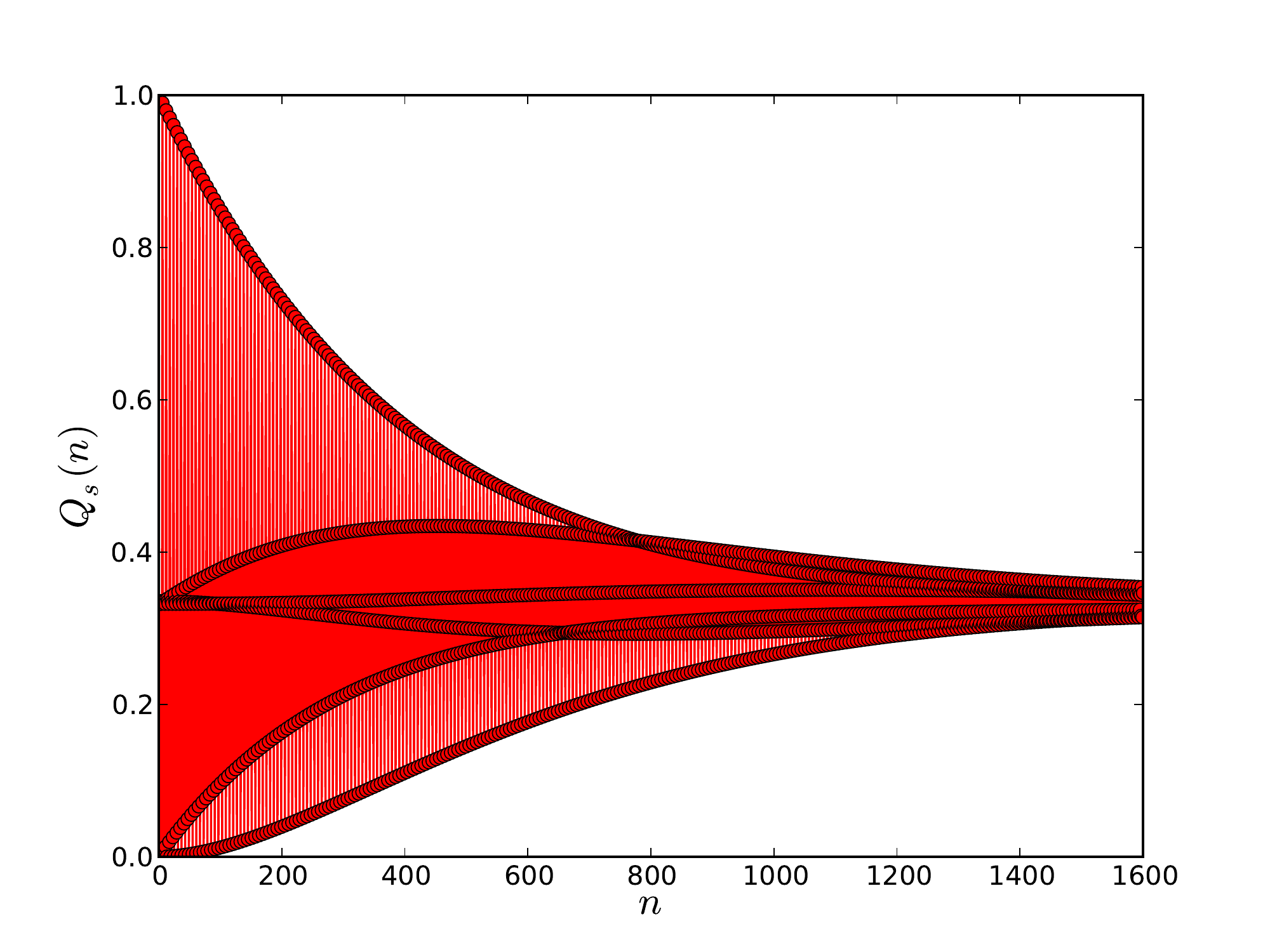}
\end{center}
\caption{$\Qsn$ vs. $n$ for the SFSF Process with $\gamma = 0.01$. This specimen is only
  very weakly faulted and, hence, there are small decay constants giving
  a slow decay to $1/3$.
  }
\label{fig:SFrankQs0p01gamma}
\end{figure}

\begin{figure}
\begin{center}
\includegraphics[width=0.5\textwidth]{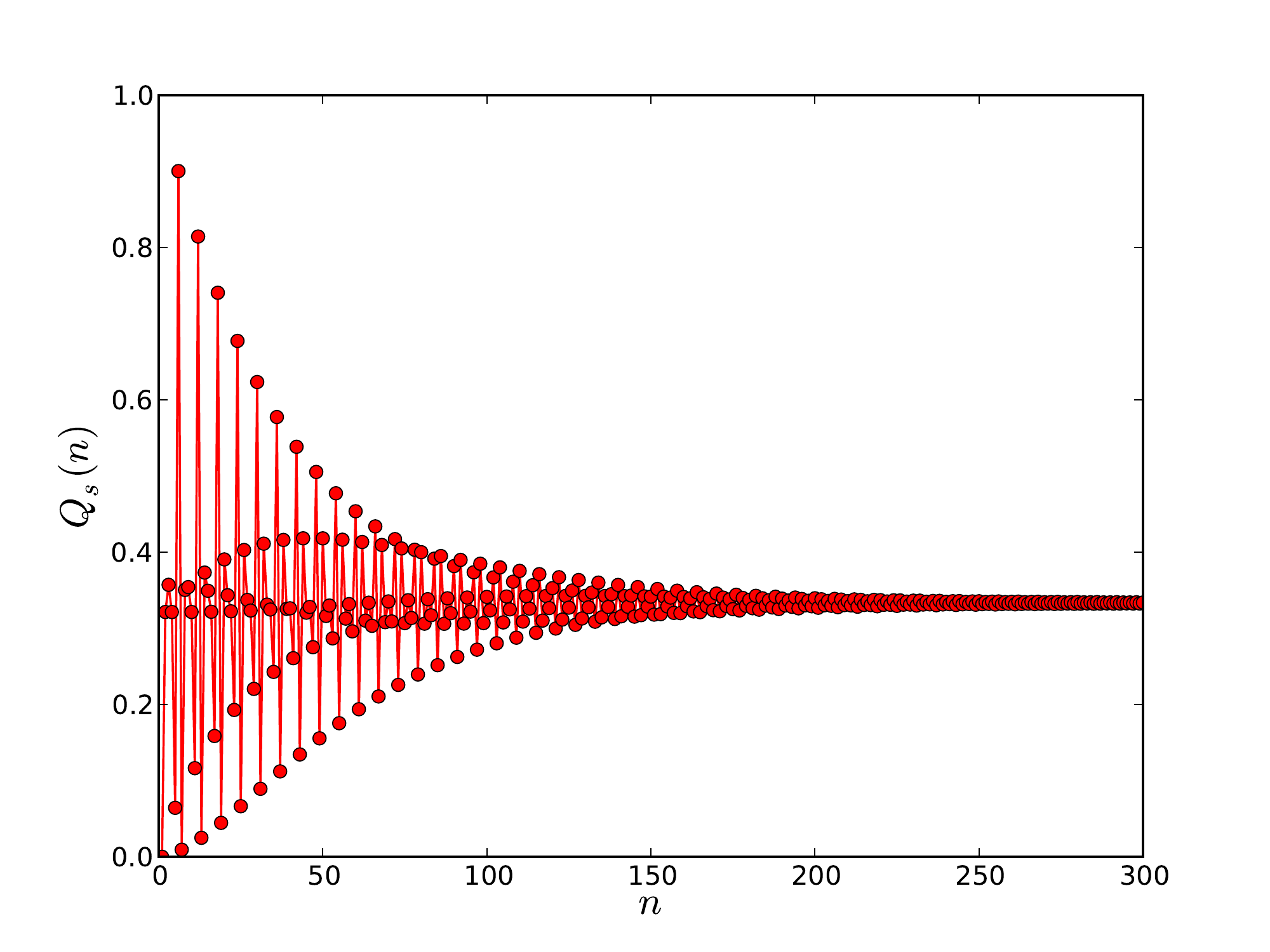}
\end{center}
\caption{$\Qsn$ vs. $n$ for the SFSF Process with $\gamma = 0.1$.
  With increasing $\gamma$, the CFs approach their asymptotic value of
  $1/3$ much more quickly.
  }
\label{fig:SFrankQs0p1gamma}
\end{figure}

\begin{figure}
\begin{center}
\includegraphics[width=0.5\textwidth]{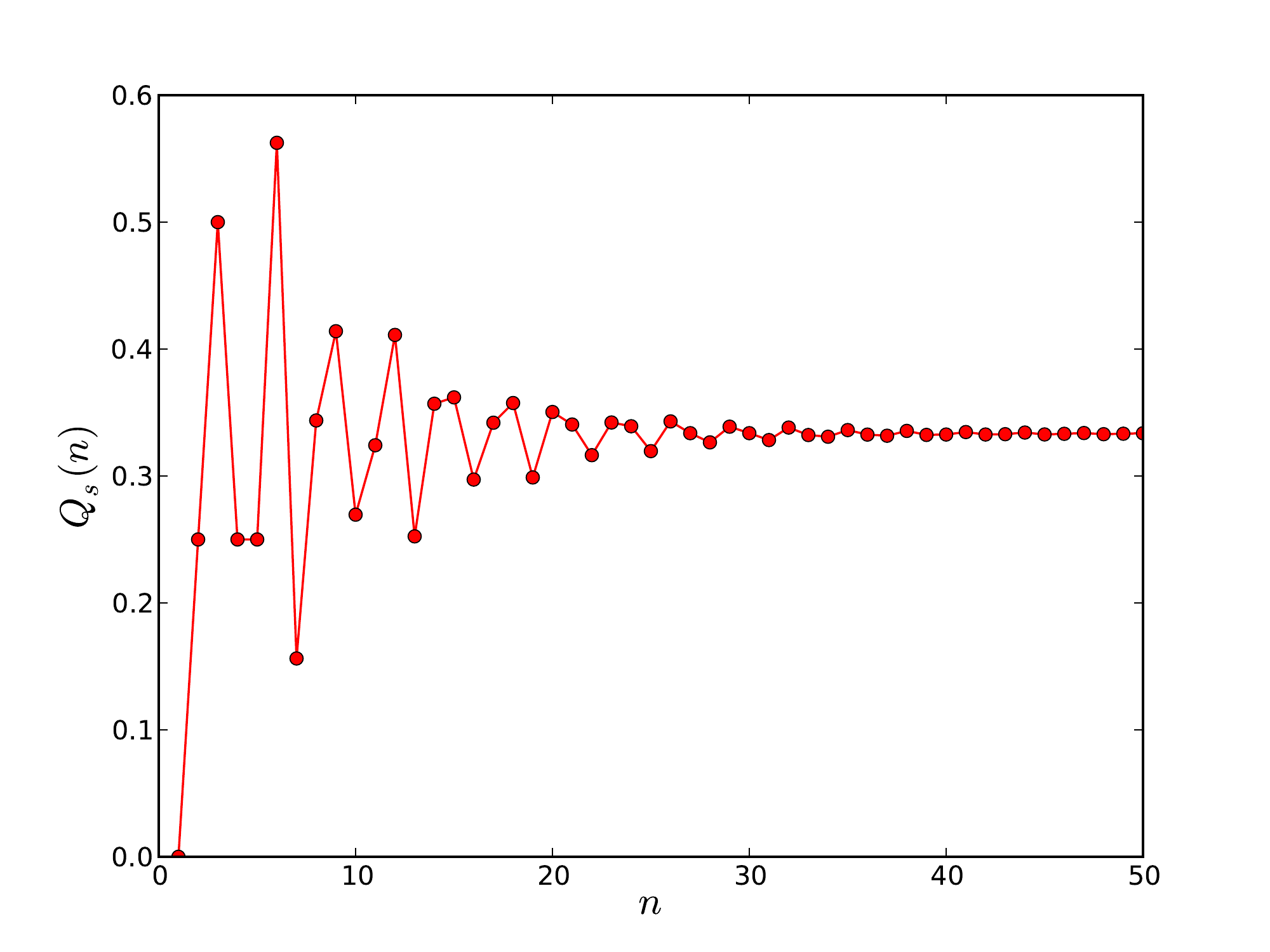}
\end{center}
\caption{$\Qsn$ vs. $n$ for the SFSF Process with $\gamma = 0.5$.
  Here, the specimen is quite disordered, and the CFs decay quickly.
  }
\label{fig:SFrankQs0p5gamma}
\end{figure}

\begin{figure}
\begin{center}
\includegraphics[width=0.5\textwidth]{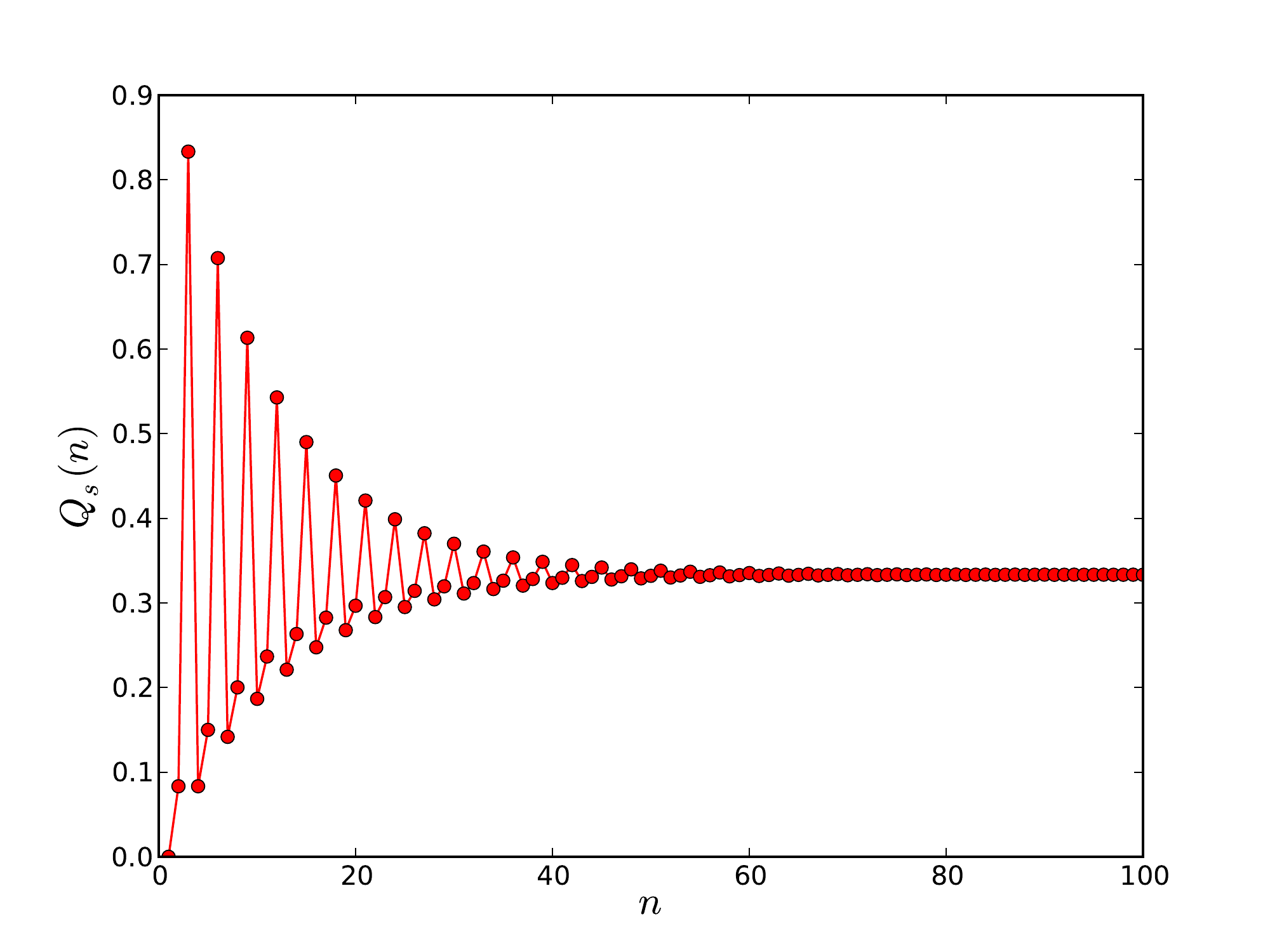}
\end{center}
\caption{$\Qsn$ vs. $n$ for the SFSF Process with $\gamma = 0.9$.
  The slower CF decay suggests that the process is now less disordered
  than the $\gamma = 0.5$ case.
  Notice that this CF is large for $n \mod(3) = 0$, indicating strong correlation
  between MLs separated by a multiple of three MLs. This is the kind of
  behavior that one expects from a twinned 3C crystal.
  }
\label{fig:SFrankQs0p9gamma}
\end{figure}

The SFSF Process's CFs are shown for several example parameter values of
$\gamma$ in
Figs.~\ref{fig:SFrankQs0p01gamma},~\ref{fig:SFrankQs0p1gamma},~\ref{fig:SFrankQs0p5gamma},
and~\ref{fig:SFrankQs0p9gamma} calculated directly from numerical
implementation of Eq.~\eqref{eq:ThreeTimesBraKetSimplification}. As the
faulting parameter is increased from $0.01 \to 0.5$, the CFs begin to decay
more quickly. However, for $\gamma = 0.9$, the correlation length increases as
the eigenvalues, near the nontrivial cube-roots of unity, loop back toward the
unit circle.  The behavior near $\gamma = 0.9$ suggests a longer ranged and
more regularly structured specimen, even though there are fewer significant
eigen-contributions to the specimen's structure.  Indeed, the bulk of the
structure is now more apparent but less sophisticated.  

%\textbf{Anything more to say? More conclusions to draw out about the particular
%cases studies? Or of the method?}

\section{Conclusion}
\label{Conclusions}

We introduced a new approach to exactly determining CFs directly from HMMs that
describe layered CPSs. The calculation can be done either with high numerical
accuracy and efficiency, as we have shown in the CF plots for each example, or
analytically, as was done for the IID and RGDF Processes. 

The mathematical representation that assumes central importance here is the
HMM. While we appreciate the value that studying CFs and, more generally, PDFs
brings to understanding material structure, pairwise
correlation information is better thought of as a consequence of a more
fundamental object (\ie, the HMM)
than one of intrinsic importance. This becomes clear when we consider that the
structure is completely contained in the very compact HMM representation. More
to the point, all of the correlation information is directly calculable from
it, as we demonstrated. In contrast, the task of inverting correlation
information to specify the underlying organization of a material's structure,
\ie, its HMM, is highly nontrivial. Over the past century considerable effort
has been expended to invert DPs, the Fourier transform of the CFs, into these
compact structural models.\footnote{We have not explicitly made the connection
here, but almost all previous models of planar disorder can generically be
expressed as HMMs.} The work of Warren~\cite{Warr69a}, Krishna and
coworkers~\cite{Seba84a,Seba87a,Seba87b,Seba87c,Seba87d}, Berliner \&
Werner~\cite{Berl86a} and that of our own
group~\cite{Varn02a,Varn07a,Varn13a,Varn13b}, to mention a few, all stand in
testament to this effort.

Although the presentation concentrated on CFs in layered CPSs, the potential
impact of the new approach is far wider. First, we note that it was necessary
to make some assumptions about the geometry of the stacking process, \ie, the
number of possible orientations of each ML and how two MLs can be stacked, in
order to demonstrate numerical results and make contact with previous work.
These assumptions however in no way limit the applicability: Any set of
stacking rules over a finite number of possible positions is amenable to this
treatment.  Second, it may seem that starting with a HMM is unnecessarily
restrictive. It is not. Given a sample of the stacking process from (say) a
simulation study, there are techniques that now have become standard for
finding the \eM, a kind of HMM, that describes the process. The subtree-merging
method~\cite{Crut89a} and causal-state splitting reconstruction~\cite{Shal02a}
are perhaps the best known, but recently a new procedure based on Bayesian
inference has been developed~\cite{Stre14a}.  Finally, a HMM may be proposed on
theoretical grounds, as done with the RGDF and SFSF HMMs in our second and
third examples. And, for the case when a DP is available, there is \eM\
spectral reconstruction theory~\cite{Varn02a,Varn07a,Varn13a,Varn13b}.  We
anticipate that HMMs will become the standard representation for describing
layered structures.

The approach presented here should also be viewed in the larger context of our
recent research thrusts. While crystallography has historically struggled to
settle on a formalism to describe disordered structures, we propose that such a
framework has been identified, at least for layered materials. Based in
computational mechanics, {\emph {chaotic crystallography}}~\cite{Varn14a}
employs information theory as a key component to characterize disordered
materials. Although the use of information theory in crystallography has been
previously proposed by Mackay and coworkers~\cite{Mack86a,Mack02a,Cart12a},
chaotic crystallography realizes this goal. Additionally, using spectral
methods in the spirit of \S \ref{subsec:SpectralResults}, information- and
computation-theoretic measures are now directly calculable from
\eMs~\cite{Crut13a,Riec14a}. And importantly, a sequel will demonstrate how
spectral methods can give both a fast and efficient method for calculating the
DP of layered CPSs or analytical expressions thereof~\cite{Riec14c}.

\section{Acknowledgment}

The authors thank the Santa Fe Institute for its hospitality during visits.
JPC is an External Faculty member there. This material is based upon work
supported by, or in part by, the U. S. Army Research Laboratory and the U. S.
Army Research Office under contract number W911NF-13-1-0390.

\bibliography{iucr}

\end{document}